\documentclass[aps,pre,twocolumn,groupedaddress,showkeys,showpacs,]{revtex4-1}

\usepackage{graphicx}
\usepackage{amsmath}
\usepackage{amssymb}
\usepackage{amsfonts}
\usepackage{subeqnarray}
\usepackage[dvips]{color}

\def\tensor#1{{\mathbf{#1}}}

\begin{document}

\title{Discrete rearranging disordered patterns:  \\
Prediction of elastic and plastic behaviour, and application to two-dimensional foams.
}

\author{C. Raufaste}
\email{christophe.raufaste@unice.fr} 
\affiliation{Laboratoire de Spectrom\'etrie Physique, BP 87, 38402 Martin d'H\`eres Cedex, France} 
\altaffiliation{UMR 5588 CNRS and Universit\'e Joseph Fourier - Grenoble I}
\affiliation{Physics of Geological Processes, University of Oslo, P.O. Box 1048, Blindern, Oslo, Norway}
\altaffiliation{Present address:  Laboratoire de physique de la mati\`ere condens\'ee, UMR 6622 CNRS and Universit\'e de Nice-Sophia Antipolis, Parc Valrose, 06108 Nice Cedex 2, France}

\author{S.J. Cox}
\affiliation{Institute of Mathematics and Physics, Aberystwyth University, SY23 3BZ, UK}

\author{P. Marmottant}
\affiliation{Laboratoire de Spectrom\'etrie Physique, BP 87, 38402 Martin d'H\`eres Cedex, France\;}

\author{F. Graner}
\affiliation{Laboratoire de Spectrom\'etrie Physique, BP 87, 38402 Martin d'H\`eres Cedex, France\;\;}

\affiliation{Institut Curie, BDD, 26 rue d'Ulm, F - 75248 Paris Cedex 05, France}
\altaffiliation{UMR 3215 CNRS, Inserm U934 and  Institut Curie}

\date{\today}

\begin{abstract}

We study the elasto-plastic behaviour of materials made of individual (discrete) objects, such as a liquid foam made of bubbles. The evolution of positions and mutual arrangements of individual objects is taken into account through statistical quantities, such as the elastic strain of the structure, the yield strain and the yield function. The past history of the sample plays no explicit role, except through its effect on these statistical quantities. They suffice to relate the discrete scale with the collective, global scale. At this global scale, the material behaves as a continuous medium; it is described with tensors such as elastic strain, stress and velocity gradient. We write the  differential equations which predict their elastic and plastic  behaviour in both the general case and the case of simple shear. An overshoot in the shear strain or shear stress is interpreted as a rotation of the deformed structure, which is a purely tensorial effect that exists only if the yield strain is at least of order 0.3. We suggest practical applications, including: when to choose a scalar formalism rather than a tensorial one; how to relax trapped stresses; and how to model materials with a low, or a high, yield strain. 

\end{abstract}

\pacs{83.80.Iz Emulsions and foams; 46.35.+z	Viscoelasticity, plasticity, viscoplasticity; 83.10.Ff  Continuum mechanics; 47.50.-d Non-Newtonian fluid flows
}

\keywords{elasticity, plasticity, overshoot, mechanics, foam, simulations}

\maketitle


\section{Introduction}
\label{sec:Introduction}

Discrete rearranging patterns include cellular patterns, for instance liquid foams, biological tissues and grains in polycrystals; assemblies of particles such as beads, granular materials, colloids, molecules and atoms; and interconnected networks \cite{Outils}. Many of these disordered  materials display elastic and plastic properties, so that  the stress tensor can rotate and is not necessarily aligned with the strain rate tensor; in models this effect is included in objective derivatives \cite{Kolymbas2003}. 

Use of simplified geometries, {\it e.g.} in a rheometer, allows a first characterization of the material through measurements of shear stress.
 An  overshoot in the shear stress  is seen during the first loading  in materials such as polymers \cite{Larson1999}, granular materials \cite{Xu2006}, and emulsions \cite{Lahtinen1988}. For liquid foams this effect has been observed in a plate-plate rheometer \cite{Khan1988} and in simulations \cite{Okuzono1995,KablaPreprintSimul}. It is unclear whether this is due to  a change in the material's structure, or a tensorial effect of shear; but nevertheless the overshoot is an essential ingredient in a recent model \cite{weaire_preprint} of the strain-rate discontinuity in the cylindrical Couette foam flow experiments of ref. \cite{lauridsen2004}. Such an overshoot results in mechanical bistability:  two different values of strain correspond to the same value of stress between the plateau and the maximum, and can thus coexist. 
 Here, we investigate the elastic regime and elasto-plastic transition in a fully tensorial model. To describe the 
mechanical behaviour we use a formalism  adapted for discrete rearranging disordered patterns
 which enables us to  quantify rotational 
effects and to test the relevant parameters   \cite{Outils}.

We use as an example  a sheared liquid foam  \cite{Abdelkader1999,Debregeas2001,Lauridsen2002,Pratt2003,Janiaud2005,Wang2006,Janiaud2006,Cheddadi2008}.
Although a liquid foam consists only of gas bubbles surrounded by liquid walls, it exhibits a complex mechanical behaviour. It is elastic for small strains, plastic for large strains and flows at large strain rates \cite{Weaire1999,Saint-Jalmes1999,Hohler2005}. This behavior is useful for numerous applications such as ore separation, oil extraction, foods and cosmetics. The individual objects, namely the bubbles, are easily identified, which makes a liquid foam (or alternatively an emulsion, made of droplets) a model for the study of other complex fluids.

This paper is organized as follows. In Section \ref{sec:Simulations}, we 
simulate the quasistatic 2D flow of a foam in a Couette shear geometry; we 
explain how we perform and represent the measurements. In Sec. 
\ref{sec:Model}, we present our equations, and discuss the specific 
effects due to the use of tensors, such as the overshoot.  Sec. 
\ref{sec:ComparisonSimulationModel} compares the model and the simulation, 
and extracts the relevant information. Sec.  \ref{sec:Practical} presents 
applications to practical situations, {\it i.e.} how and when to use the 
model. Sec.  \ref{sec:Conclusions} summarizes our findings. An Appendix 
explains the notation and provides the detailed equations. 

\section{Simulations}
\label{sec:Simulations}

We simulate numerically a   2D foam flowing  in a linear Couette shear geometry. Simulations of dry foams offer several advantages: (i) the parameters are homogeneous (liquid fraction, bubble area) and controlled (no diffusion-driven coarsening or film rupture);  (ii) the yield strain is of order of 0.3, which is large enough to observe a full tensorial elastic regime while small enough that plastic effects can be easily observed; (iii) all physical quantities can be easily measured.

\begin{table}
\centering
\begin{tabular}{|c||c|c|c|c|}
\hline
symbol & $\Phi_{\rm eff}$ & $\delta A/A$ & geometry& $\gamma_{max}$\\
\hline
\hline
{\large \color{black}$\blacktriangledown$} or {\large \color{black}$\triangledown$} & $9.7.10^{-5}$ & 0 & fully periodic & $\pm$ 2\\
\hline
{\color{blue}${\mathbf{\times}}$} or {\color{blue}${\mathbf{+}}$}  & $3.9.10^{-4}$ & 0 & fully periodic & $\pm$ 2\\
\hline
{\color{green}$\blacklozenge$}  or {\color{green}$\lozenge$} & $3.9.10^{-4}$ & 0.025 & fully periodic & $\pm$ 2\\
\hline
{\color{cyan}$\blacktriangle$} or {\color{cyan}$\triangle$} & $3.9.10^{-4}$ & 0.66 & fully periodic & $\pm$ 2\\
\hline
{\Large \color{red}$\bullet$}  or {\Large \color{red}$\circ$} & $3.5.10^{-4}$ & 0 & confined & $\pm$ 2.5\\
\hline
{\color{magenta}$\blacksquare$} or {\color{magenta}$\square$} & $3.5.10^{-4}$ & 0.66 & confined & $\pm$ 2.5\\
\hline
\end{tabular}
\caption{
Characteristics of simulated foams. The different columns correspond to the symbols used in Figs. 
\ref{Fig:PlaPlasticLimit} and \ref{Fig:PlaOvershootUs}, effective liquid fraction \cite{Raufaste2007}, area dispersity,  boundary conditions and maximal amplitude of the cycles. 
}\label{Tab:DefSimul}
\end{table}

\begin{figure}[!h]
\setlength{\unitlength}{1cm} 
\centering
\begin{picture}(8,12)(0.0,0.0)
\put(0,8.1){\includegraphics[width=4cm]{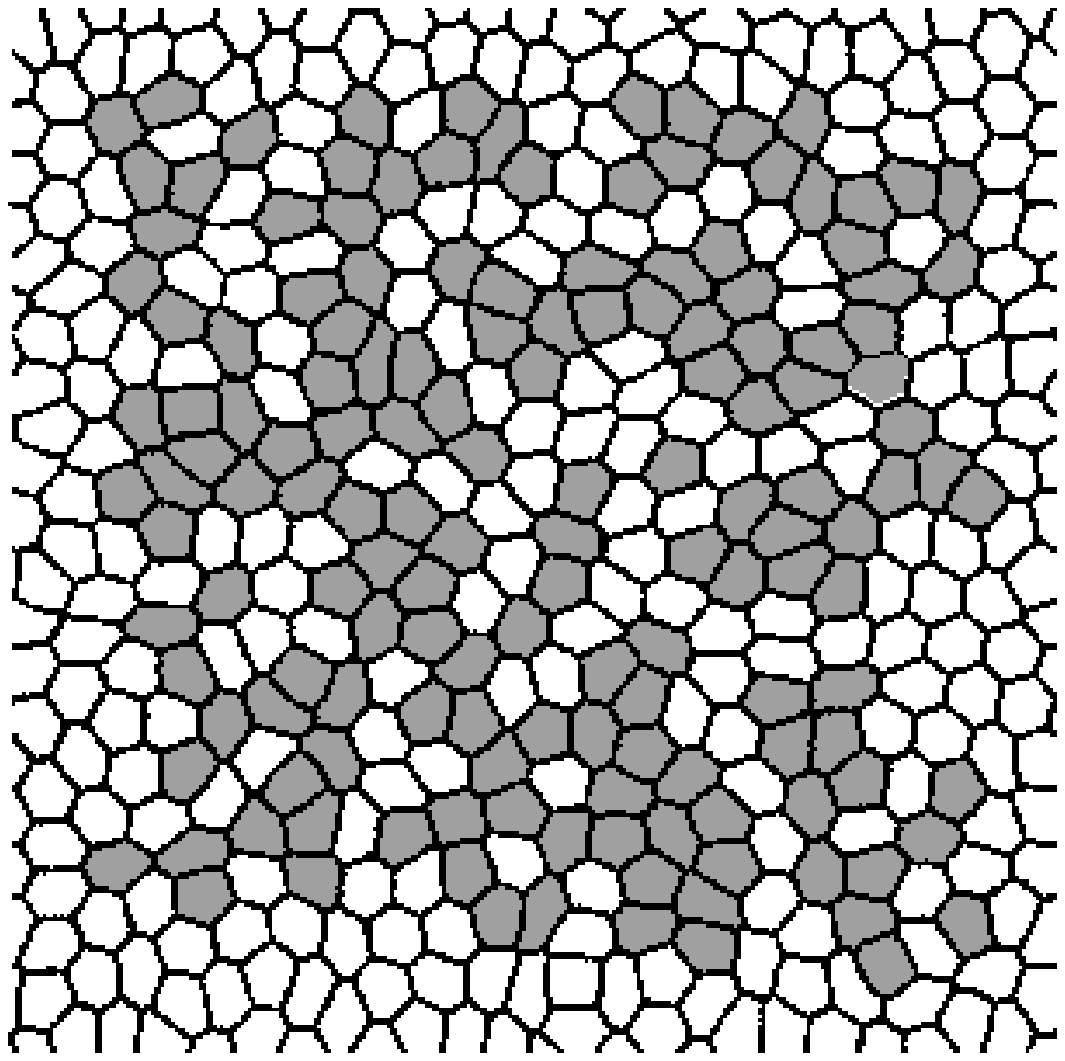}}
\put(4.5,8.1){\includegraphics[width=4cm]{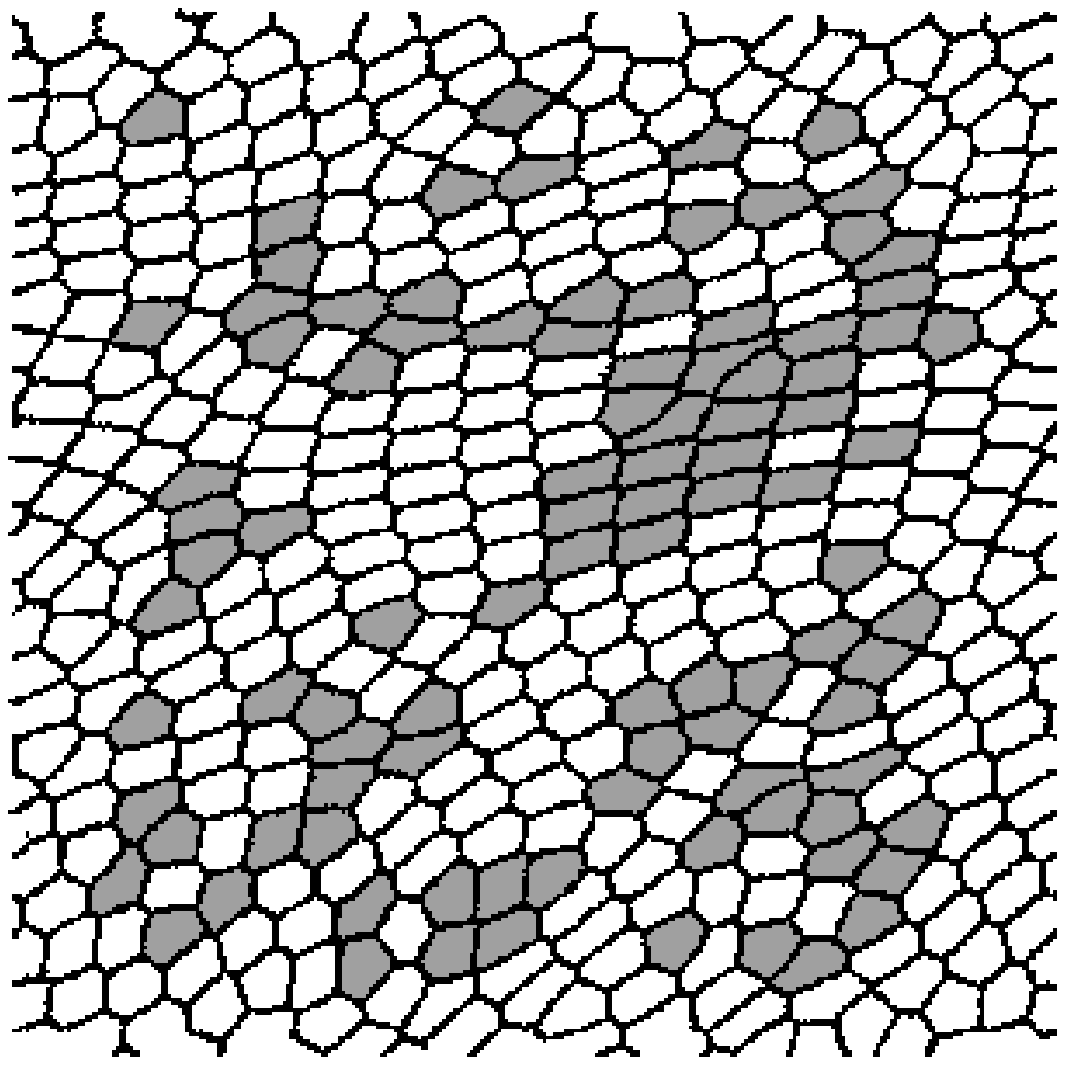}}
\put(0,4){\includegraphics[width=4cm]{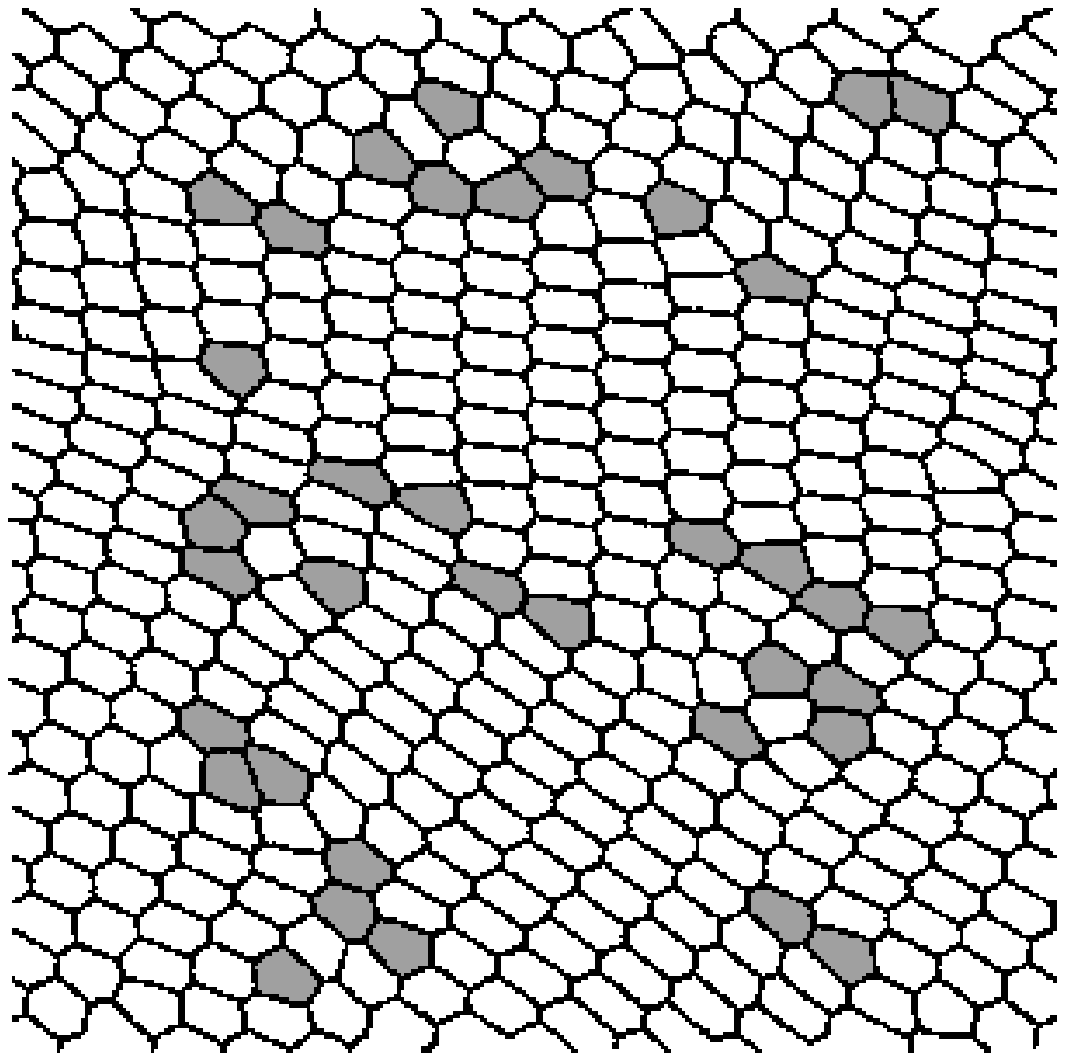}}
\put(4.5,4){\includegraphics[width=4cm]{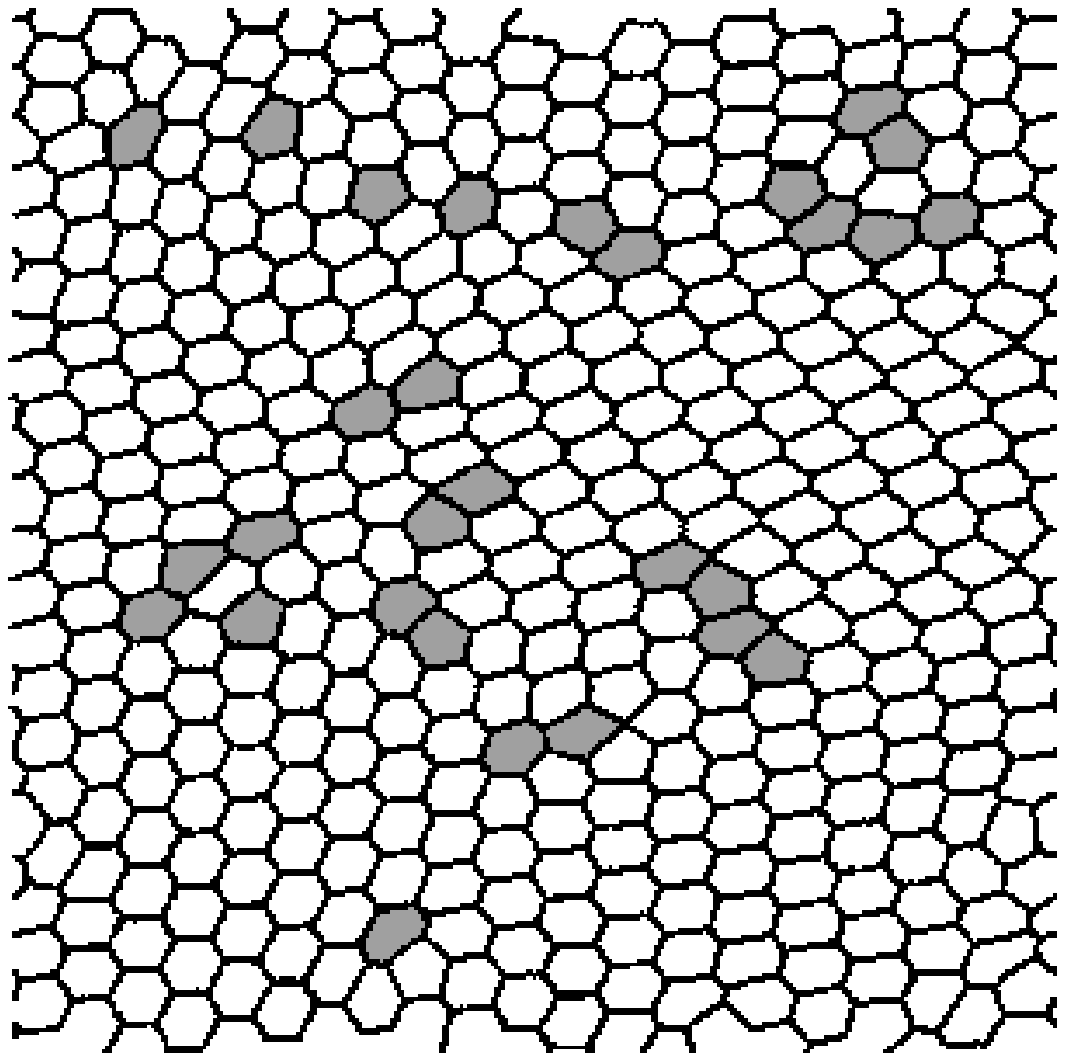}}
\put(2,-0.1){\includegraphics[width=4cm]{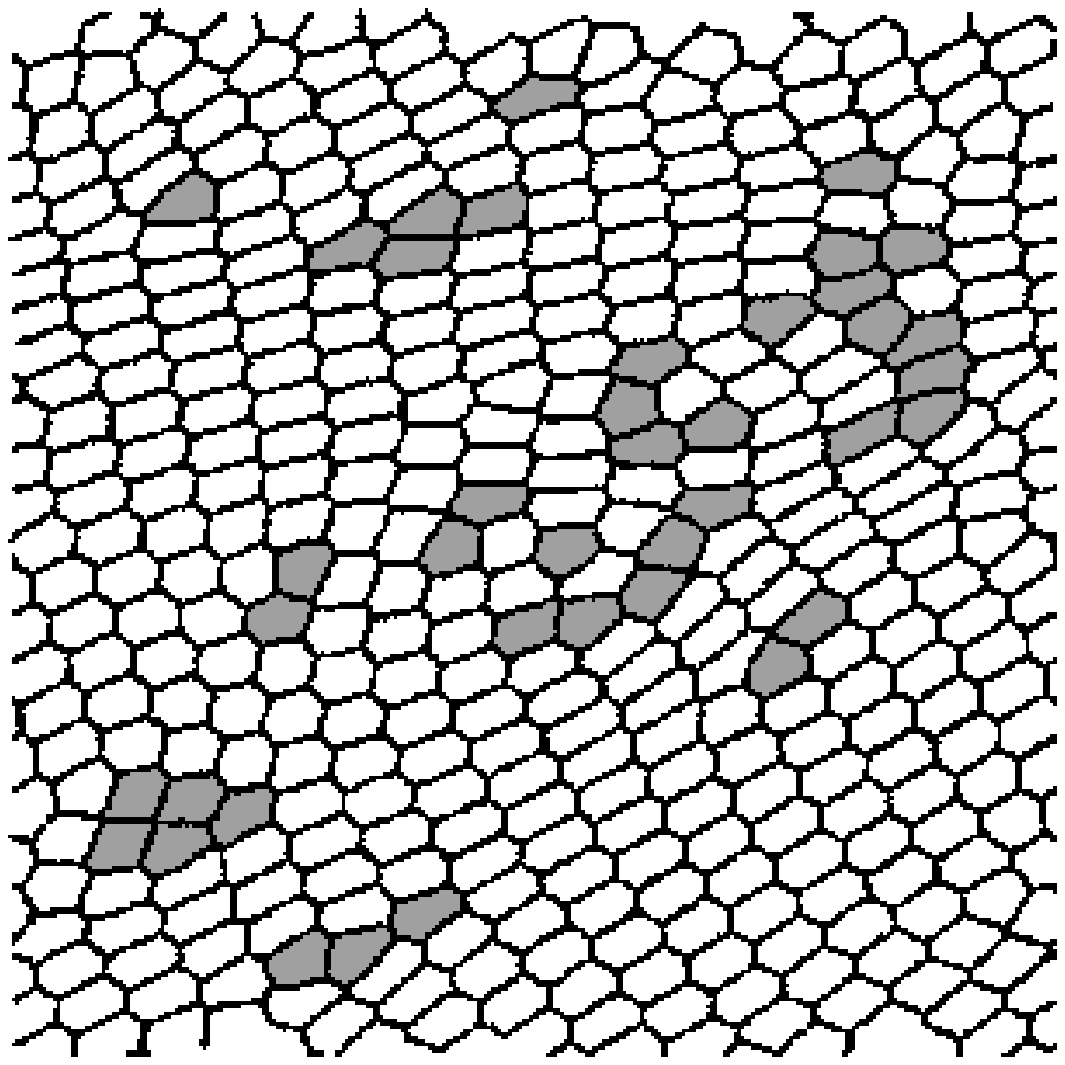}}
\put(-0.5,8.1){\large{\textbf{(1)}}}
\put(4,8.1){\large{\textbf{(2)}}}
\put(-0.5,4){\large{\textbf{(3)}}}
\put(4,4){\large{\textbf{(4)}}}
\put(1.5,-0.1){\large{\textbf{(5)}}}
\end{picture}
\caption{
Example of 2D foam simulation. Pictures are successive snapshots of a  quasi-statically sheared, fully periodic   foam. Numbers correspond  to those of Figs. \ref{Fig:integral_strain} and \ref{Fig:SimulationRepresentation}. Bubbles with 6 neighbours are displayed in white, otherwise in gray.}
\label{Fig:PlaStructureRefFP}
\end{figure}

\subsection{Methods}
\label{sec:NumericalApproach}

Several ideal, two-dimensional, dry foams \cite{Weaire1999} are simulated 
(Fig. \ref{Fig:PlaStructureRefFP} and Table \ref{Tab:DefSimul}). 
We use the Surface Evolver \cite{brakke92} in a mode in which each film is
represented as a circular arc. The value of surface tension is taken equal to 1 throughout, without loss of generality. A realistic foam structure is found by minimizing the total film length
subject to the constraint of fixed bubble areas, prescribed at the beginning of the simulation.
 The simulations are quasistatic, which means that the system has time to relax between successive time steps (increments in applied strain). Relaxation effects are thus neglected and viscosity does not need to be included. The behaviour is expected to be elasto-plastic. 
 
The simulation procedure is as follows. A Voronoi construction of randomly distributed points
\cite{brakke86}  (not shown) is first used to generate a fully periodic tessellation of
the plane. To create a confined foam, bubbles at the top and bottom
are sequentially deleted until the required number of bubbles remains.
In each case, the structure is imported into the Surface Evolver and target bubble areas prescribed, either all the same (monodisperse, $\delta A/A = 0$), a small random variation of up to 20\% about monodisperse ($\delta A/A = 0.025$), or equal to the areas given by the Voronoi construction ($\delta A/A = 0.66$).

The initial foam configuration for each simulation ({\it e.g.} label (1) in Fig. \ref{Fig:PlaStructureRefFP}) is found by reducing the total
film length to a local minimum. During this minimization, neighbour swappings (so-called ``T1s" \cite{Weaire1999}) are
triggered by deleting each film that shrinks below a certain critical length $l_c$ and allowing a new film to form to complete the process. The critical length $l_c$ defines and measures  an effective liquid fraction, $\Phi_{\rm eff}$ \cite{Raufaste2007}, here chosen to be very dry (Table \ref{Tab:DefSimul}).

One geometry consists of a unit cell of 400 bubbles with fully periodic boundary conditions to eliminate any artefacts due to small sample sizes. The second geometry mimics more closely a real experiment, and consists of 296 bubbles with two parallel bars (about 15 bubble diameters apart) confining the foam and with periodicity in one direction only.

To shear the foams, two different procedures are required. For the periodic foams, one off-diagonal component of the matrix describing the periodicity of the unit cell is adjusted by a small amount \cite{brakke92}.  For the confined foams, a small step in strain is applied by moving one of the
confining walls a small distance and then moving all vertices affinely. In each case this is followed by reduction of the film length to a minimum correct to 16 d.p. using conjugate gradient (without biasing the search by introducing any large-scale perturbations of the structure).

\begin{figure}[!h]
\includegraphics[width=8cm]{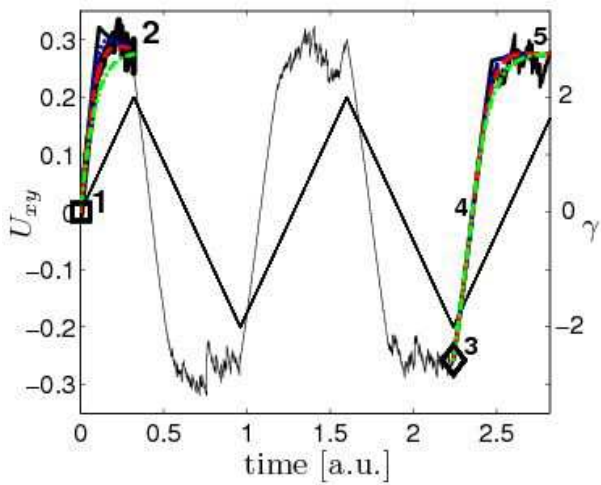}
\caption{
Time evolution of the reference 
simulation ({\color{blue}$\times$} in Tab. \ref{Tab:DefSimul}). Horizontal axis: time is in arbitrary units, equivalent to the ``cumulated strain" $ \int | \dot{\gamma}|dt$, where $t$ is the time and $ \dot{\gamma}$ is defined up to an arbitrary prefactor; here 2.25 cycles are represented. Vertical axis: all curves represent $U_{xy}$ (left scale) except for the saw-tooth which is the applied $\gamma$ (right scale).  Numbers correspond to the pictures in Fig. \ref{Fig:PlaStructureRefFP}. The first step, plotted with a thick solid line, starts at the  $\Box$ (indicated also by a number 1) and its end is labelled by number 2: $\gamma=0\to 2$.  The second step, plotted with a thin solid line,  is from number 2 to 3. The third step, plotted with a middle solid line, starts at the $\lozenge $  (indicated also by a number 3), and extends to number 5: $\gamma=-2\to 2$.  Four predictions of the model are plotted as dashed lines (see Fig. \ref{Fig:Pla-h-n} for explanation of the legend); for clarity they are plotted only from 1 to 2 and from 3 to 5: note that from 3 to 4 all predictions are indiscernable from the simulation.
}
\label{Fig:integral_strain}
\end{figure}

 Each foam is subjected to at least two ``saw-tooth" shear cycles of amplitude $\gamma_{max}$. Positive and negative steps correspond respectively to shear toward increasing or decreasing imposed strain $\gamma$ (Fig. \ref{Fig:integral_strain})

\begin{figure}[!h]
\includegraphics[width=8cm]{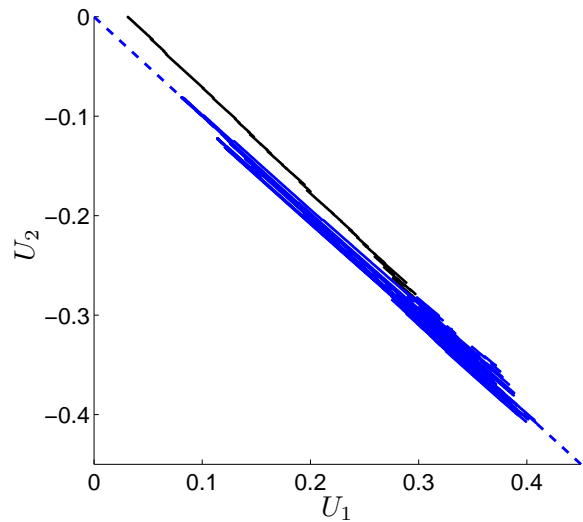}\\
{\large{\textbf{a)}}}\\
\includegraphics[width=8cm]{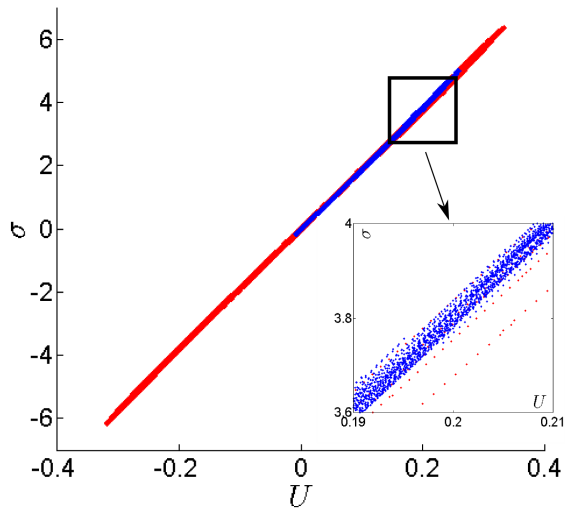}\\
{\large{\textbf{b)}}}
\caption{Validation of hypotheses. Both approximations of section 
\ref{sec:Measurements} are tested during the shear of the reference 
simulation ({\color{blue}$\times$} in Tab. \ref{Tab:DefSimul}).
 a) 
 Strain eigenvalues $U_1$ {\it vs} $U_2$;  black solid line: initial 
shear ($1-2$)which anneals the disorder; blue dots: shear cycles ($2-5$); dashed blue line: straight line of slope -1 passing through the 
origin.
 b) 
Deviatoric stress-strain relation: red, $\sigma_{xy}$ {\it vs} 
$U_{xy}$; blue, $(\sigma_{xx}- \sigma_{yy})/2$ {\it vs} $(U_{xx}- 
U_{yy})/2$; red and blue data almost perfectly overlap (see zoom in 
inset); the slope determines $2\mu$, where $\mu$ is the shear modulus.
}
\label{Fig:VerifUTraceLess}
\end{figure}

\subsection{Measurements}
\label{sec:Measurements}

 At each step, the positions of the bubble centres and films are recorded.
Tensorial quantities  are measured  by averaging over all bubbles, as follows  \cite{Outils}.

The texture tensor $\tensor{M}=\langle \vec{\ell} \otimes \vec{\ell} \rangle$  is computed statistically as
an average over  vector links $\vec{\ell}$ between  centres of neighbouring bubbles. We assume here that the reference texture at rest,  $\tensor{M}_0$, is isotropic. We thus define it by measuring the average of  the determinant of $\tensor{M}$ over the duration of the whole simulation,
$\det(\tensor{M}_0) = \left\langle \det(\tensor{M}) \right\rangle$.
 The elastic strain  of bubbles expresses the deviation from the reference state,   $ \tensor{U}
  =  \left(  \log \tensor{M} - \log   \tensor{M}_0 \right)/2$ (Eq. \ref{Eq:Fundament_4_ann}).
  
  This tensor is symmetric by construction ($U_{yx} = U_{xy}$); it can be diagonalized and has two eigenvalues ($U_{1}$, $U_{2}$) in two orthogonal eigendirections.
The simulations   (Fig. \ref{Fig:VerifUTraceLess}a) verify that we can reasonably assume its trace to be always close to zero, $U_1+U_2 = U_{xx}+U_{yy}\approx 0$, as is roughly expected  for  an incompressible material \cite{Outils}. Thus, due to its symmetry and vanishing trace, $\tensor{U}$ has only two independent components:
\begin{equation}\label{Eq:StressMatrix}
\tensor{U} = 
\left(
\begin{array}{cc}
U_{xx} & U_{xy}\\
U_{yx} & U_{yy}
  \end{array}
\right)
\approx
\left(
\begin{array}{cc}
\frac{U_{xx}-U_{yy}}{2} & U_{xy}\\
U_{xy} & -\frac{U_{xx}-U_{yy}}{2}
  \end{array}
\right)
.
\end{equation}
Fig. \ref{Fig:VerifUTraceLess}a shows that our measurement of the elastic strain makes evident the effect of shear-induced shuffling  \cite{{QuillietPreprint}}: the annealed foam (blue dots) differs significantly from the initial one (black solid line). 

 The contribution to the stress of the network of bubble walls is obtained by integrating over all films 
   \cite{bachelor,coxwhittick}; it yields the 
  deviatoric ({\it i.e}. traceless) part of the elastic stress tensor  $\tensor{\sigma}$. 
   The trace of the stress, namely the pressure, is unimportant here. The simulations are quasistatic and the viscous part of the stress is not relevant. 
  
We check (Fig. \ref{Fig:VerifUTraceLess}b) that the stress and strain are strongly correlated; that their correlation is linear; and that it is isotropic (the same for $xy$ and $xx-yy$ components) \cite{Asipauskas2003,Cartes}. Half the slope thus defines and measures the elastic shear modulus $\mu$.

\begin{figure}[!h]
\setlength{\unitlength}{1cm} 
\centering
\begin{picture}(7,4)(0.0,0.0)
\put(-1,0){\resizebox{4.cm}{!}{\includegraphics{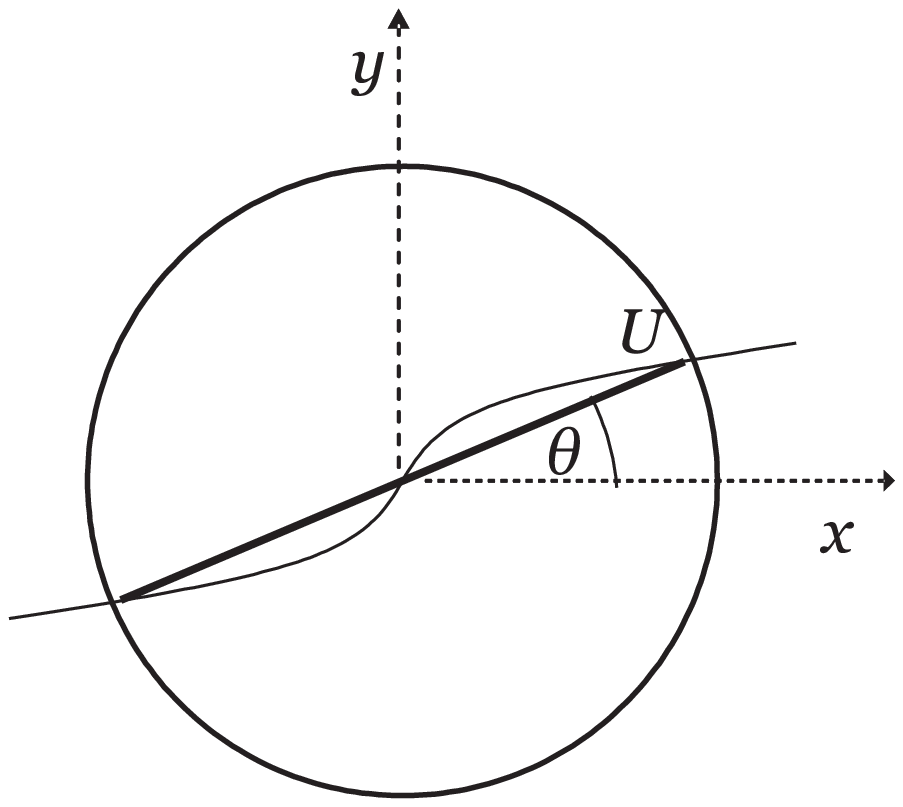}}}
\put(3.5,0){\resizebox{4.3cm}{!}{\includegraphics{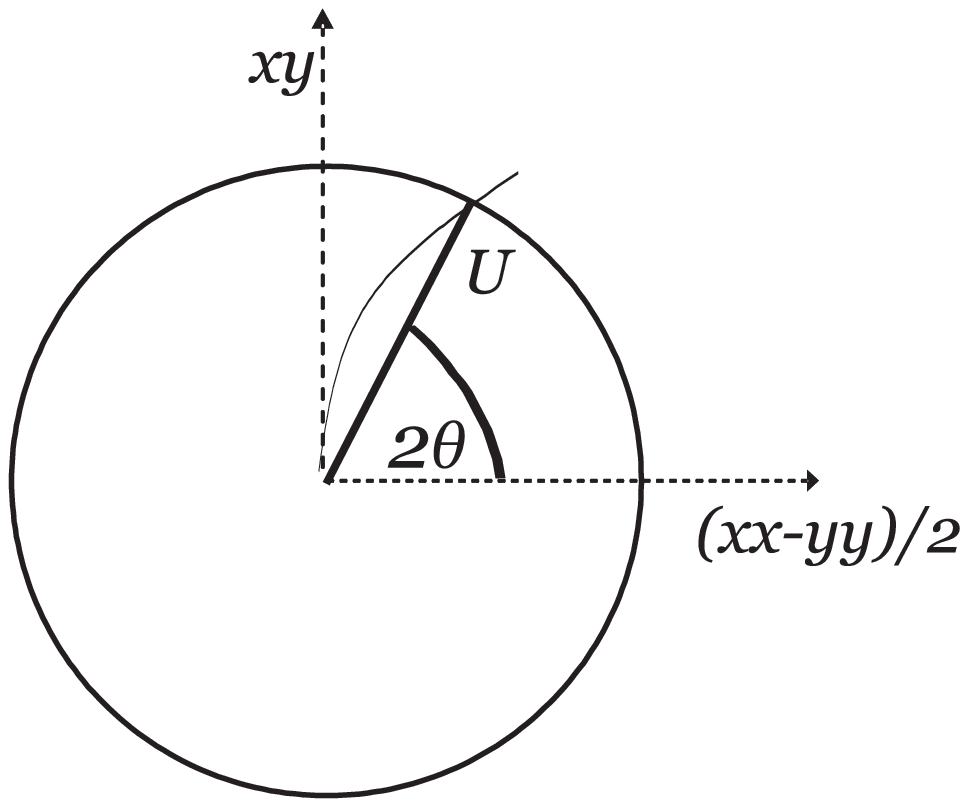}}}
\put(-0.5,0){\large{\textbf{a)}}}
\put(3.5,0){\large{\textbf{b)}}}
\end{picture}
\caption{Representation of the evolution of $\stackrel{=}{U}$. a) \emph{Physical space}: evolution of the point ($U\cos{\theta},U\sin{\theta}$); for completeness we also plot the opposite (and strictly equivalent) point 
($-U\cos{\theta},-U\sin{\theta}$). b) \emph{Component space}: trajectory of ($(U_{xx}-U_{yy})/2,U_{xy}$), that is, ($U\cos{2 \theta},U\sin{2 \theta}$).}
\label{Fig:RepresentationPhysicalStress}
\end{figure}

\subsection{Representations}
\label{sec:Representations}

To summarize,  $\tensor{M}$,  $\tensor{U}$ and  $\tensor{\sigma}$ characterize 
  the current state of the foam.
 These three tensors carry here the same information, since $\mu$ appears constant. 
In what follows, texture, elastic strain and stress tensors are always aligned.

We choose to display $\tensor{U}$ only, because it is dimensionless, and thus more general: it makes the comparison of different materials easy. 
One possibility \cite{Outils} is to represent the traceless tensor  $\tensor{U}$ as a circle of radius $U$, with a straight line to indicate  the direction $\theta$ of its positive eigenvalue: see thick lines (circle and straight line) in Fig. \ref{Fig:RepresentationPhysicalStress}a. We do not use it here, except in the inset of Fig. \ref{Fig:PlaPlasticLimit}. In fact, it is easier to represent $\tensor{U}$ at a given time by a point, enabling us to plot trajectories. 
Its two independent components can be represented in two different but equally useful ways, as follows.
  Both representations are equally appropriate in the problem considered here because of the circular symmetry of the yield criterion (see Eq. \ref{Eq:UCriterion}). 
    
 First, in the case of a traceless tensor,  the absolute value of the two eigenvalues is the same and equal to   the amplitude $U$ of the tensor $\tensor{U}$ defined as
\begin{equation}\label{Eq:UPrincipalComponent_3}
  U  = \sqrt{ \left( U_{xx}-U_{yy} / 2 \right)^2 + U_{xy}^2},
\end{equation}
or equivalently 
 $U  = \left|  (U_{1}-U_{2})/2\right| =  \left|   \left|   \tensor{U} \right| \right| / \sqrt{2}$, where  $  \left|   \left|  \tensor{U} \right| \right|
= \left(\Sigma_{ij} \left(U_{ij}\right)^2\right)^{1/2}$ is the euclidian norm of $\tensor{U}$.
We call $\theta$ the direction of the greatest eigenvalue.
We call \emph{physical space} the representation of the parameters ($U$,$\theta$).
It is useful because it shows the evolution of the structure (elongation, orientation), 
In particular, we plot the trajectory of the point ($U \cos \theta(\gamma), U \sin \theta(\gamma)$) (Fig. \ref{Fig:RepresentationPhysicalStress}a). 
 
The other representation, which has already been used for foams \cite{KablaPreprintSimul}, is called \emph{component space}. It plots the trajectory of the point ($(U_{xx}(\gamma)-U_{yy}(\gamma))/2, U_{xy}(\gamma)$) (Fig. \ref{Fig:RepresentationPhysicalStress}b).
It is more suitable for comparison with experimental data, since rheometers measure the tangential 
stress ($xy$), and sometimes the normal stress difference  ($xx-yy$). 

\begin{figure}[!h]
\setlength{\unitlength}{1cm} 
\centering
\begin{picture}(14,13)(0.0,0.0)
\put(-1.7,0){\resizebox{10cm}{!}{\includegraphics{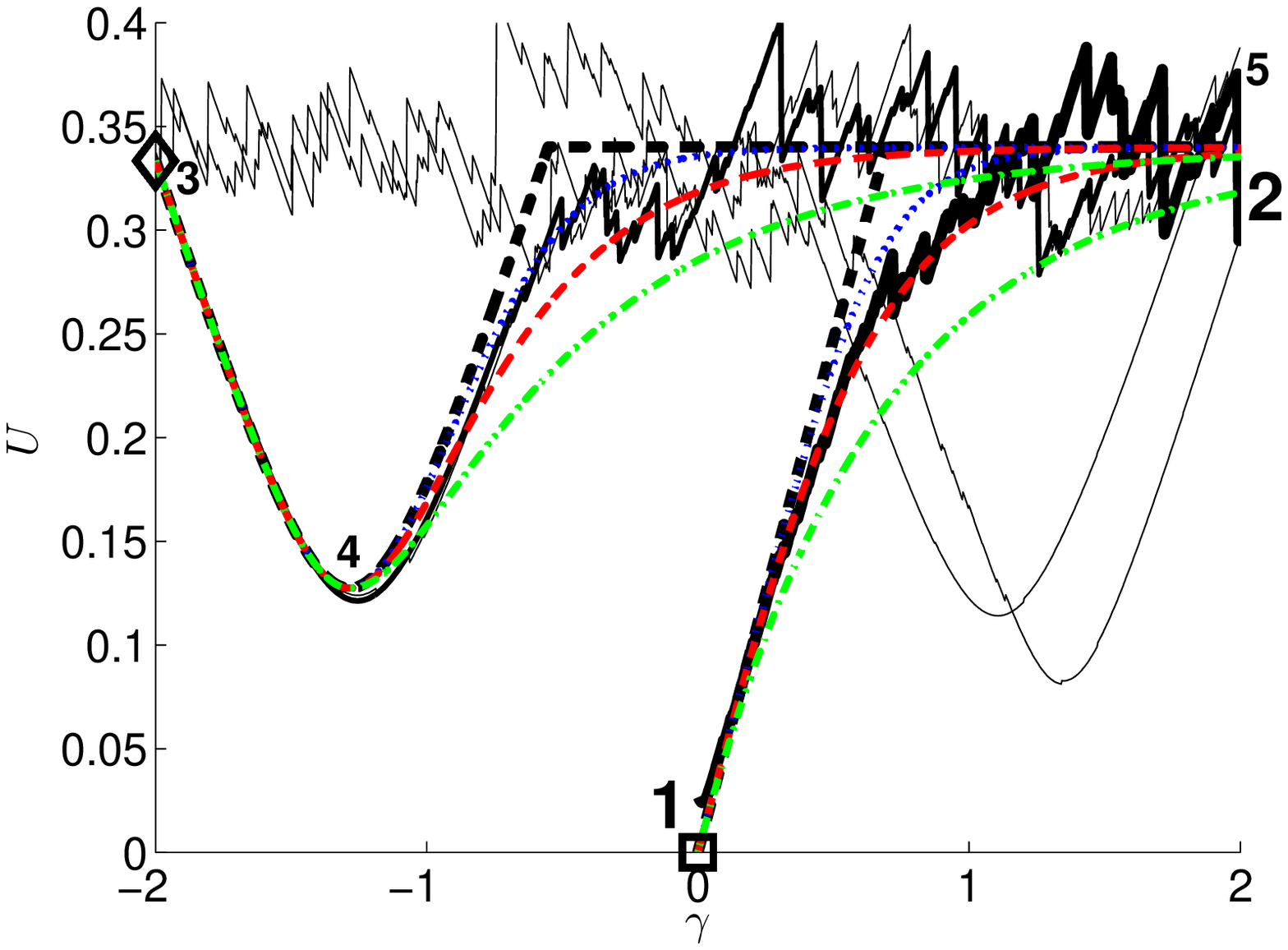}}}
\put(-1.2,0){\large{\textbf{c)}}}
\put(-1.7,7.1){\resizebox{10cm}{!}{\includegraphics{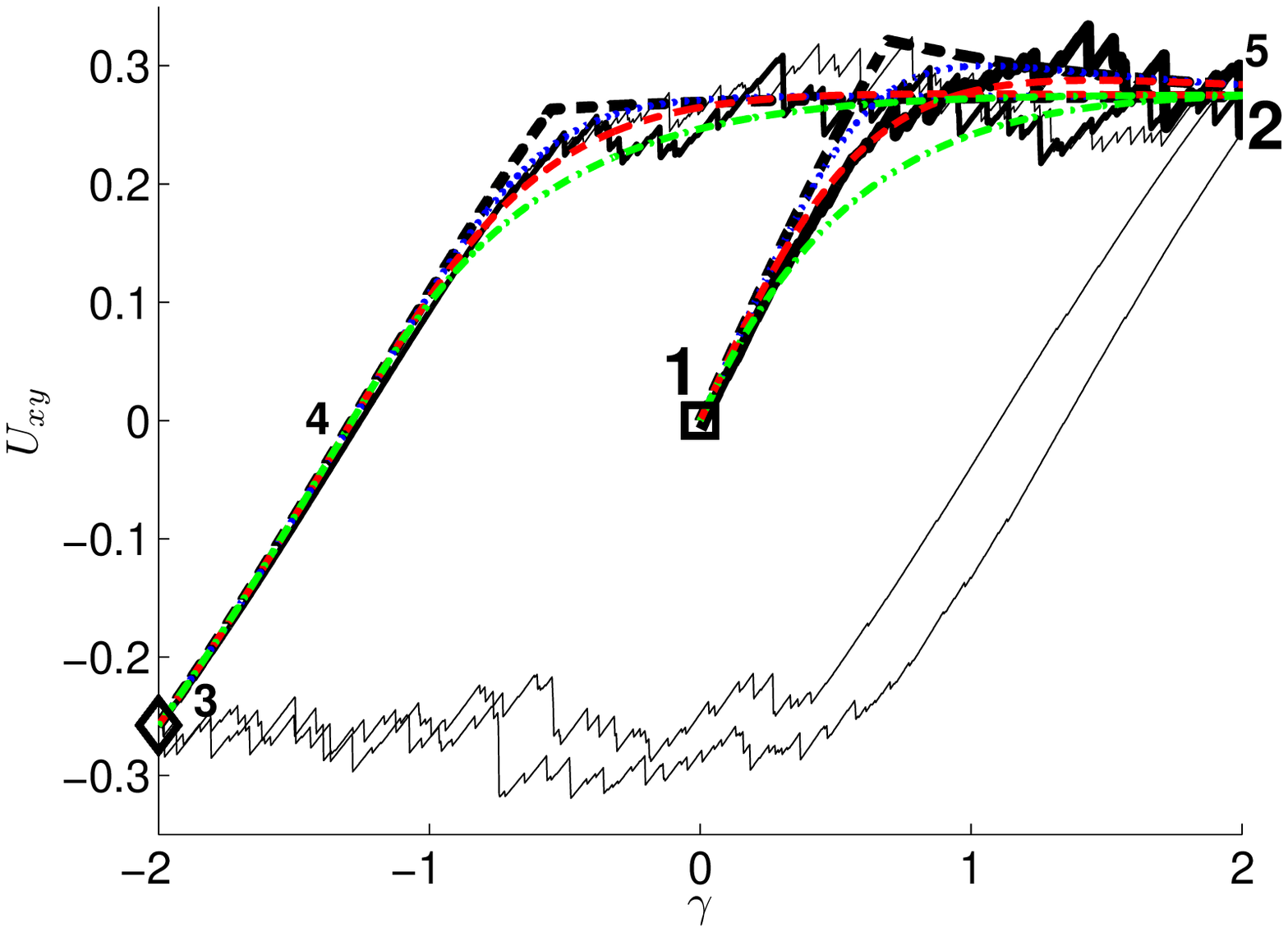}}}
\put(-1.2,7.1){\large{\textbf{a)}}}
\put(8.4,0){\resizebox{7.7cm}{!}{\includegraphics{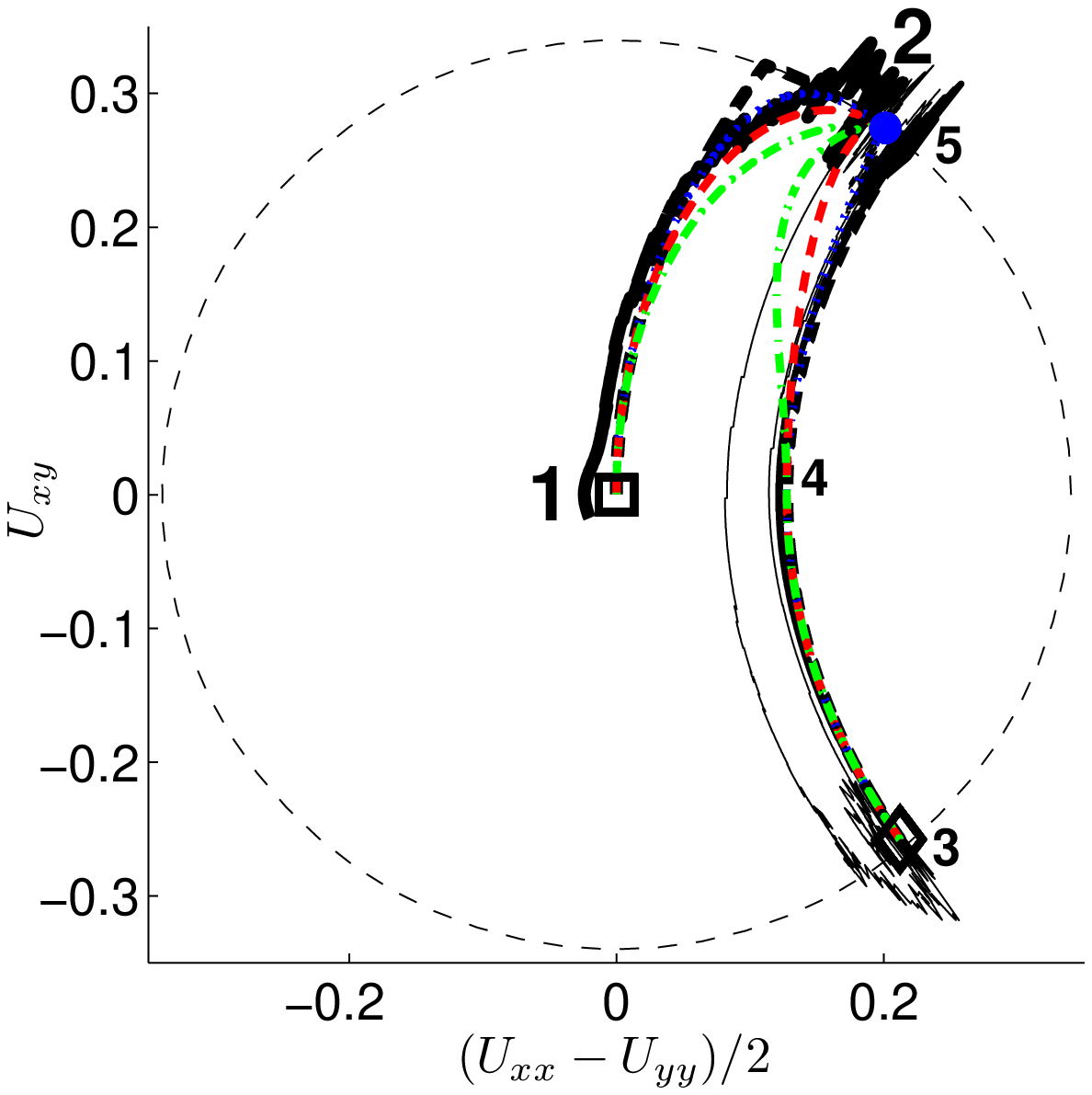}}}
\put(8.5,0){\large{\textbf{d)}}}
\put(8.1,7){\resizebox{7.7cm}{!}{\includegraphics{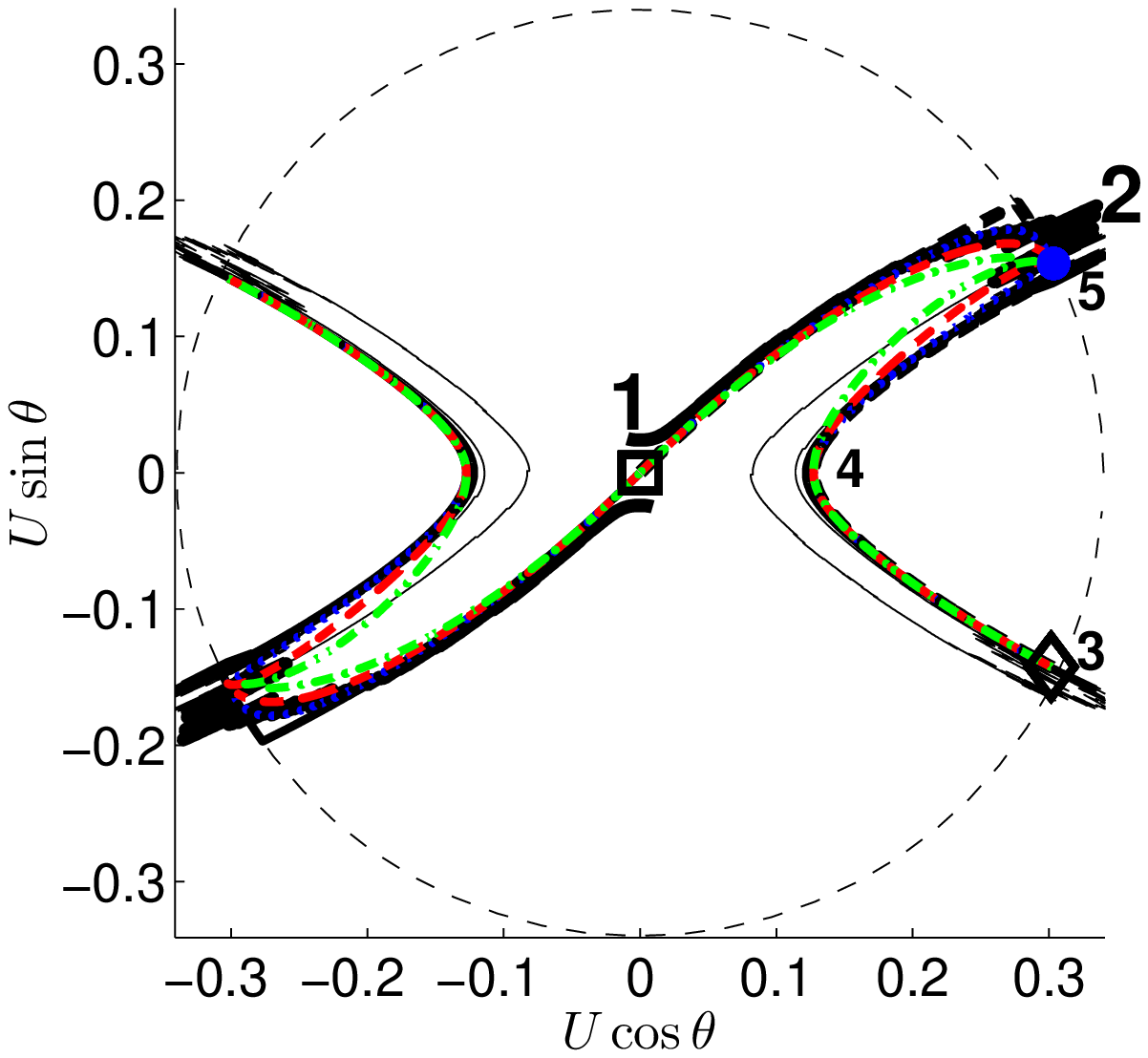}}}
\put(8.5,7.1){\large{\textbf{b)}}}
\end{picture}
\caption{Different representations of Fig. \ref{Fig:integral_strain} (same symbols) according to  Fig. \ref{Fig:RepresentationPhysicalStress}.
a) $U_{xy}$ versus $\gamma$.
b) \emph{Physical space}. c) $U$ versus $\gamma$. d) \emph{component space}.
The dashed circle in b) and d) has radius $U_Y = 0.34$. 
 }
\label{Fig:SimulationRepresentation}
\end{figure}

These two possible choices are related by
\begin{subeqnarray}
\label{Eq:UPrincipalComponent}
\slabel{Eq:UPrincipalComponent_1}
 & U_{xy} & =  U\sin{2\theta} \\
\slabel{Eq:UPrincipalComponent_2}
 & \displaystyle  \frac{U_{xx}-U_{yy}}{2} & =  U\cos{2\theta}.
\end{subeqnarray}
Complete data for one simulation are plotted in Fig. \ref{Fig:SimulationRepresentation} and are discussed in the next section.

\newpage

\onecolumngrid

\twocolumngrid

\section{Model}
\label{sec:Model}

\subsection{Implementation}
\subsubsection{Elasticity equations}
\label{sec:Formulation}

 As  already mentioned, we consider a quasistatic limit in which flow is slow enough that we may neglect viscous stresses. The stress is then related to the elastic strain (Fig \ref{Fig:VerifUTraceLess}b). We don't consider here the effect of external forces, such as gravity or friction on the boundary if the system is confined between glass plates \cite{Wang2006}.
 
In the present 2D case, classical plasticity \cite{Hill1950} suggests that the material begins to yield when the difference between the stress eigenvalues becomes too large: $(\sigma_{1}-\sigma_{2})^2 = 4 \sigma_{Y}^2$.
  The yield stress  $\sigma_Y$  separates a domain of pure elasticity from a domain in which the material flows plastically. A complete set of continuous equations (Reuss equations \cite{Hill1950}) can then be derived;  $\sigma_{Y}$ is assumed to be constant (no strain hardening). The effect of pressure (trace of the stress) is  neglected, which is usually a good first approximation for metals for instance \cite{Hill1950}. 
  It must be even more appropriate for soft materials, like a foam, for which the shear modulus is several orders of magnitude smaller than the bulk modulus \cite{Weaire1999}. In what follows we prefer to use the component yield criterion \cite{Hill1950}:
\begin{equation}\label{Eq:ComponentYieldCriterion}
\left(\frac{\sigma_{xx}-\sigma_{yy}}{2}\right)^2 + \sigma_{xy}^2 = \sigma_{Y}^2 .
\end{equation}

Equivalently, Eq. \ref{Eq:ComponentYieldCriterion} can be  written for  $\tensor{U}$, since the deviatoric parts of   $\tensor{\sigma}$ and $\tensor{U}$ are proportional (Fig. \ref{Fig:VerifUTraceLess}b). From Eqs. \ref{Eq:UPrincipalComponent_3} and \ref{Eq:ComponentYieldCriterion}, we can write the yield criterion  as
\begin{equation}\label{Eq:UCriterion}
U = U_{Y}.
\end{equation}
This is represented by a circle in both physical and component spaces (Fig. \ref{Fig:SimulationRepresentation}bd)
  
\begin{figure}[!h]
\setlength{\unitlength}{1cm} 
\centering
\begin{picture}(8,5)(0.0,0.0)
\put(-0.4,-0.1){\includegraphics[height=5cm]{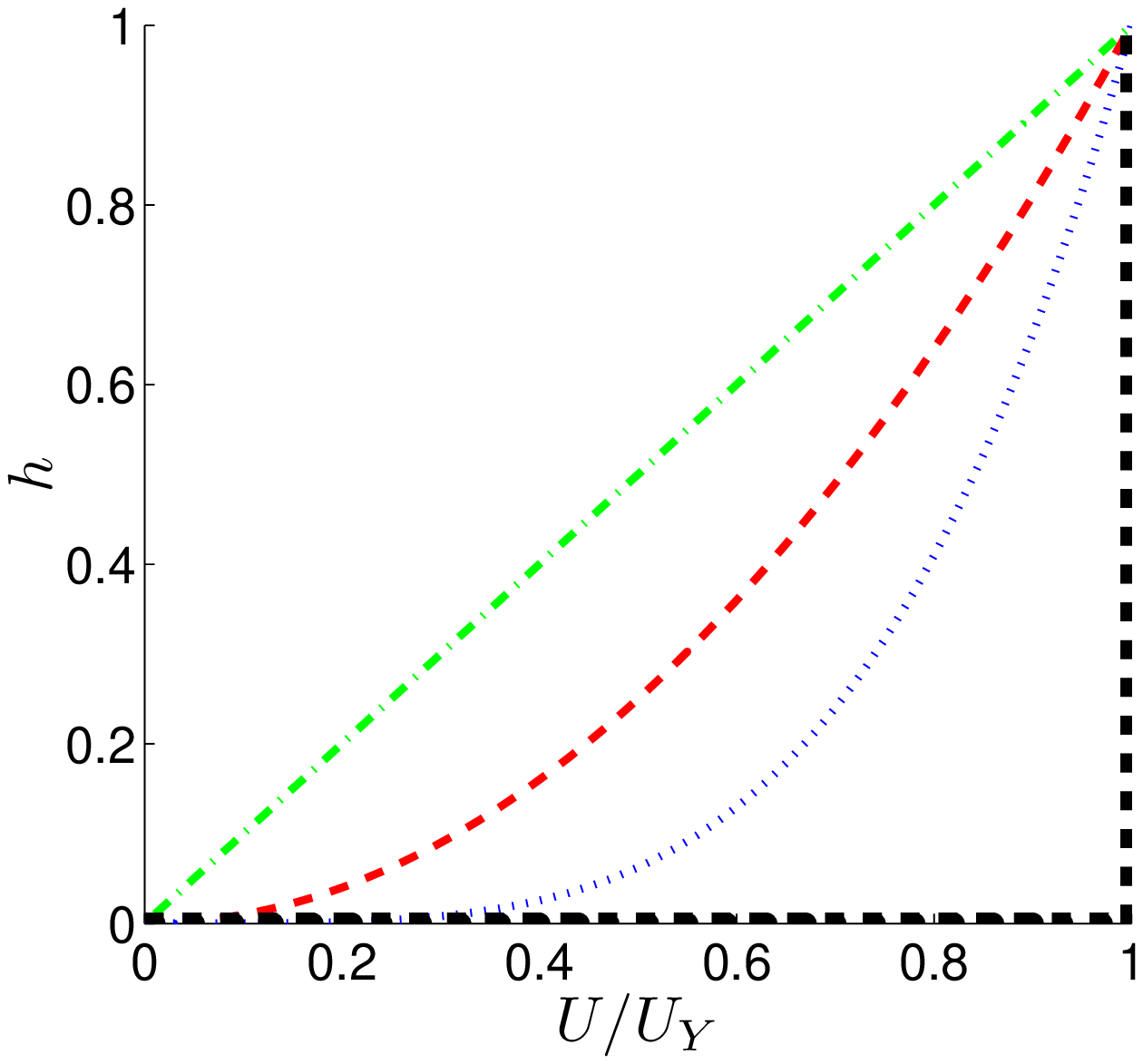}}
\put(5.8,0){\includegraphics[height=5cm]{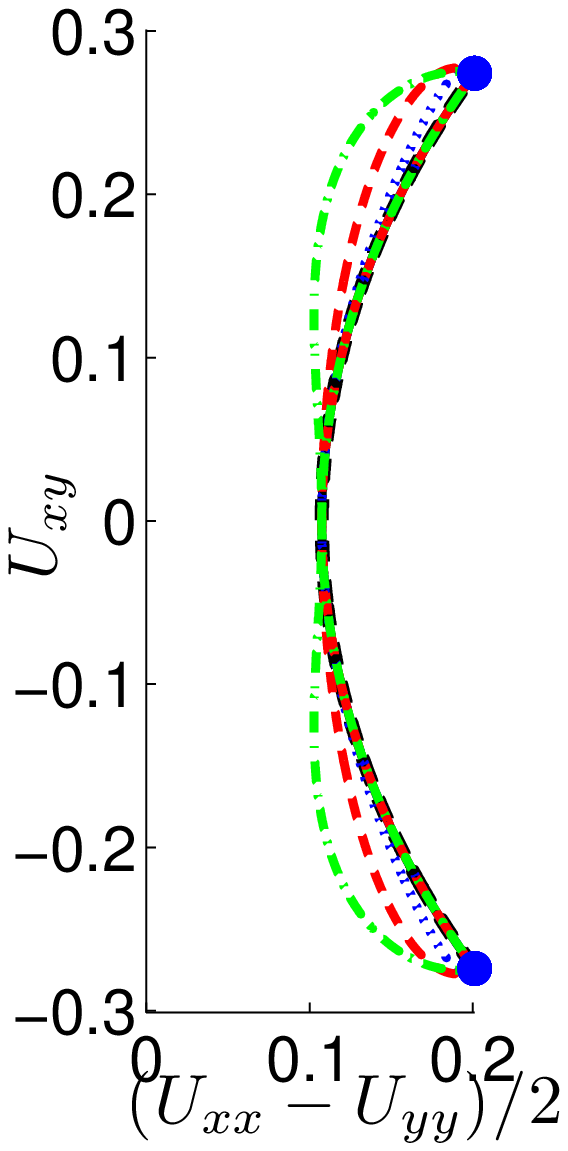}}
\put(0,0){\large{\textbf{a)}}}
\put(5.8,0){\large{\textbf{b)}}}
\end{picture}
\caption{(color online) Model. (a)  The different yield functions $h$ used as examples in the present paper are power laws $h(U)= (U/U_Y)^n$ with $n=1$ (green, dash-dots), $n=2$ (red, dashes), $n=4$ (blue, dots) and $n=+\infty$ (black, thick dashes, equal to 0 everywhere except at $U=U_Y$ where it is equal to 1).
b) Corresponding  limit cycles predicted by the model, plotted in
\emph{component space} (same legend).
}
\label{Fig:Pla-h-n}
\end{figure}

For the example of foams and highly-concentrated emulsions, Marmottant and Graner \cite{MarmottantPinceau} suggested that the transition between elastic and plastic regimes is not sharp, but can be described by a yield function $h$. This function is 0, respectively 1, in the pure elastic, respectively plastic, domain. Between these two limits, both effects are present and the proportion seems to depend mostly on the elastic strain amplitude $U$ (Fig. \ref{Fig:Pla-h-n}a). This assumption was successfully tested on different flow geometries of a 2D foam  \cite{Cartes}. 

 By assuming that the deformation is affine \cite{Outils} and according to the prediction of plasticity  (Eq. 22 in \cite{Cartes}), we can then write a  tensorial equation of evolution of the texture  (Eq. 9 in \cite{Outils}). This dictates the evolution of the texture due to imposed strain (deformation, rotation) and due to relaxation (rearrangements), for details see Eqs. \ref{Eq:Fundament_3_ann},\ref{Eq:PlasticMarmottant}: 
\begin{eqnarray}\label{Eq:EvolutionEquation}
	\frac{d}{d t}  \tensor{M} & = & \tensor{M}.\tensor{\nabla v} + \tensor{\nabla v}^{t}.\tensor{M} \nonumber \\
	& & - \left(\frac{\tensor{U}}{U}:\tensor{\nabla v}_{sym}\right)\;  {\cal H} \;  h\left(\frac{U}{U_{Y}}\right) \frac{\tensor{U}}{U} .\tensor{M} \nonumber\\
	& & 
\end{eqnarray}
Here  $d/dt$ is the Lagrangian derivative in time (including advection); $ \tensor{M}.\tensor{\nabla v} + \tensor{\nabla v}^{t}.\tensor{M} $ is the variation of $\tensor{M}$ due to convection by the velocity gradient $\tensor{\nabla v}$; as explained in appendix \ref{sec:NotationsForTensors} and \ref{sec:TensorComponents}, the notation 
 ${U}:\tensor{\nabla v}_{sym}$ is the scalar product  of   the elastic strain tensor
 with the symmetrized velocity gradient tensor
 $\tensor{\nabla v}_{sym}= \left(\tensor{\nabla v} + \tensor{\nabla v}^{t}\right)/2$
  \cite{Outils}; conversely, $\tensor{U} .\tensor{M} $ is the usual product of tensors; here
  $  {\cal H} =  {\cal H}\left(\frac{\tensor{U}}{U}:\tensor{\nabla v}_{sym}\right)$ is the Heaviside function, which is equal to 1 if  ${U}:\tensor{\nabla v}_{sym}$ is positive and 0 otherwise. 
  
Eq. \ref{Eq:EvolutionEquation} links the evolution of the foam texture with the elastic strain. It is quasistatic  in the sense that the strain is relevant, not the strain-rate.  It can  be generalized to evolutions quicker than the relaxation times of the structure \cite{Saramito2007}. 
  Plasticity   occurs only when the elastic strain is oriented in the direction of shear, as expressed by the Heaviside function ${\cal H}$ (Appendix \ref{sec:ElastoPlasticComponentEquations}).  
  
 The model is continuous and analytic, without fluctuations. The information regarding disorder is recorded in $h$. 
Trapped stresses   \cite{Larson1999} are recorded in the initial value $\tensor{M}_i$  (or equivalently $\tensor{U}_i$).
The material's yielding criterion is encoded in $U_Y$. The history of the material only plays a role in determining  $h$, $\tensor{M}_i$  (or $\tensor{U}_i$), and $U_Y$, 
which together fully describe the material.  According to the expression of  $h$, Eq. \ref{Eq:EvolutionEquation} can be integrated analytically or numerically.

\subsubsection{Simple shear}
\label{sec:Implementation}

To study the structure-evolution equation, {\it i.e.} the competition between elasticity and plasticity, and predict the rheological behaviour, we impose a strain rate $\dot{\gamma}$ on the material.
We take $x$ as the direction of the shear, which gives the following velocity field:
\begin{equation}
	\tensor{\nabla v}=
	\left(
\begin{array}{cc}
\partial_{x} v_{x} & \partial_{x} v_{y} \\
\partial_{y} v_{x} & \partial_{y} v_{y}
\end{array}
\right)
= \dot{\gamma}
\left(
\begin{array}{cc}
0 & 0 \\
1 & 0
\end{array}
\right)\label{Eq:PlaSimplificationGradV}
\end{equation}
and hence
\begin{equation}
	\tensor{\nabla v}_{sym}
=  \dot{\gamma}
\left(
\begin{array}{cc}
0 & 1/2 \\
1/2 & 0
\end{array}
\right).\label{Eq:PlaSimplificationGradVsym}
\end{equation}
This factor $1/2$ appears when comparing the scalar and tensorial descriptions 
(appendix \ref{sec:AnnScalarLimit}). 
In this geometry, the advection term is taken equal to zero and the resulting system of equations is given in Appendix \ref{sec:ElastoPlasticComponentEquations}.
 The reference state $\tensor{M}_0$ is considered isotropic and constant throughout the evolution.

We recall that this evolution is quasistatic: $\dot{\gamma}$ appears as a prefactor in the time evolution, Eqs. \ref{Eq:Ann2EquationQuasistatique}. We thus follow the evolution with the strain $\gamma=\int  \dot{\gamma}\; dt$,  instead of the time.
Tensor operations and the time evolution of $\tensor{M}$ are implemented by a finite difference procedure.
Between two time steps, $\tensor{U}$ is recalculated with Eq. \ref{Eq:Fundament_4_ann}.

\subsection{Predictions}
\label{sec:Predictions}

We now address the resolution of the full elasto-plastic set of equations \ref{Eq:Ann2EquationQuasistatique}. 
Our representation underlines the specifically tensorial effects.

\subsubsection{Purely elastic regime}
\label{sec:elasticRegime}

\begin{figure}[!h]
\setlength{\unitlength}{1cm} 
\centering
\begin{picture}(8,4)(0.0,0.0)
\put(-0.5,0){\includegraphics[width=4.6cm]{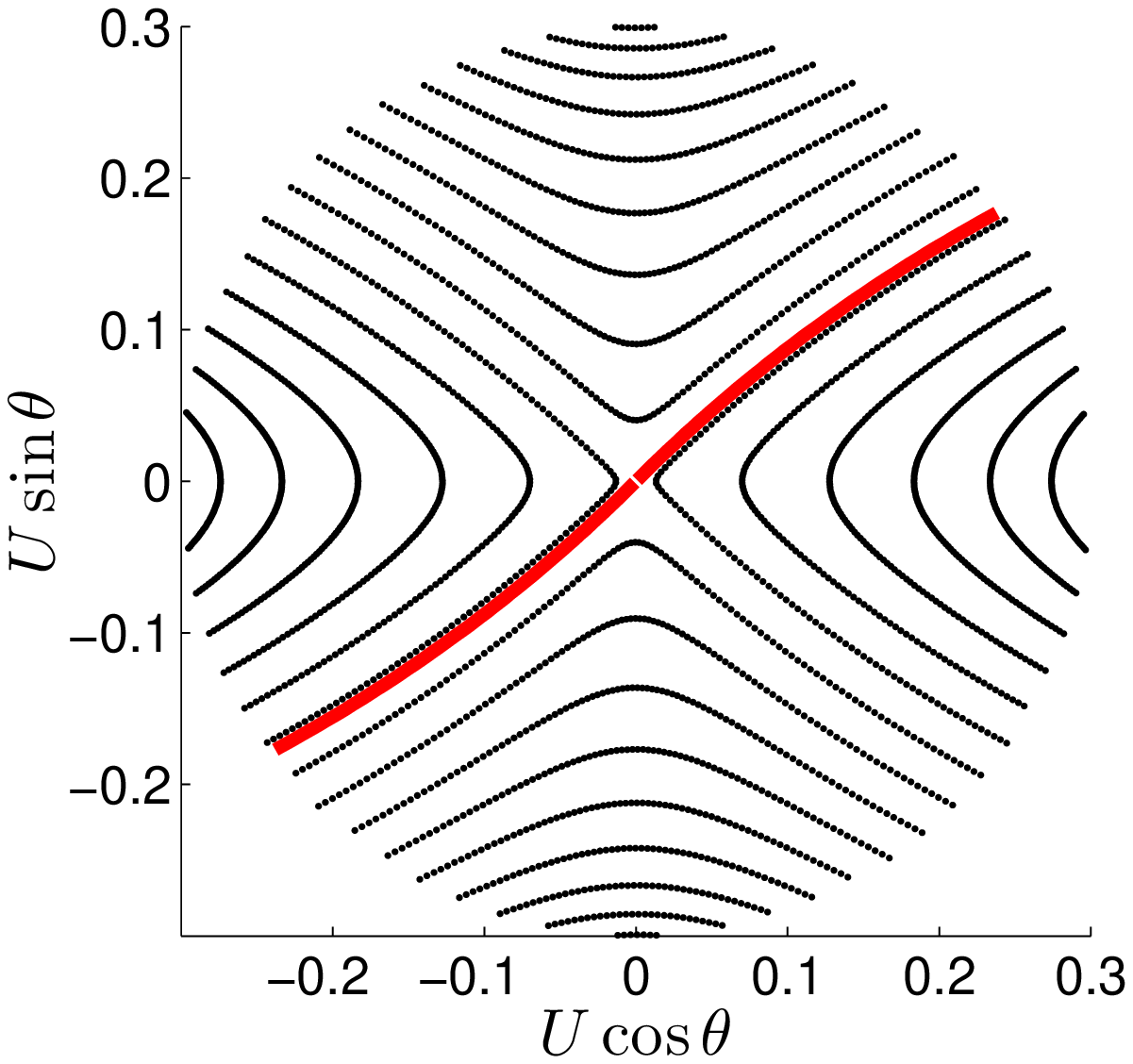}}
\put(3.8,0){\includegraphics[width=4.9cm]{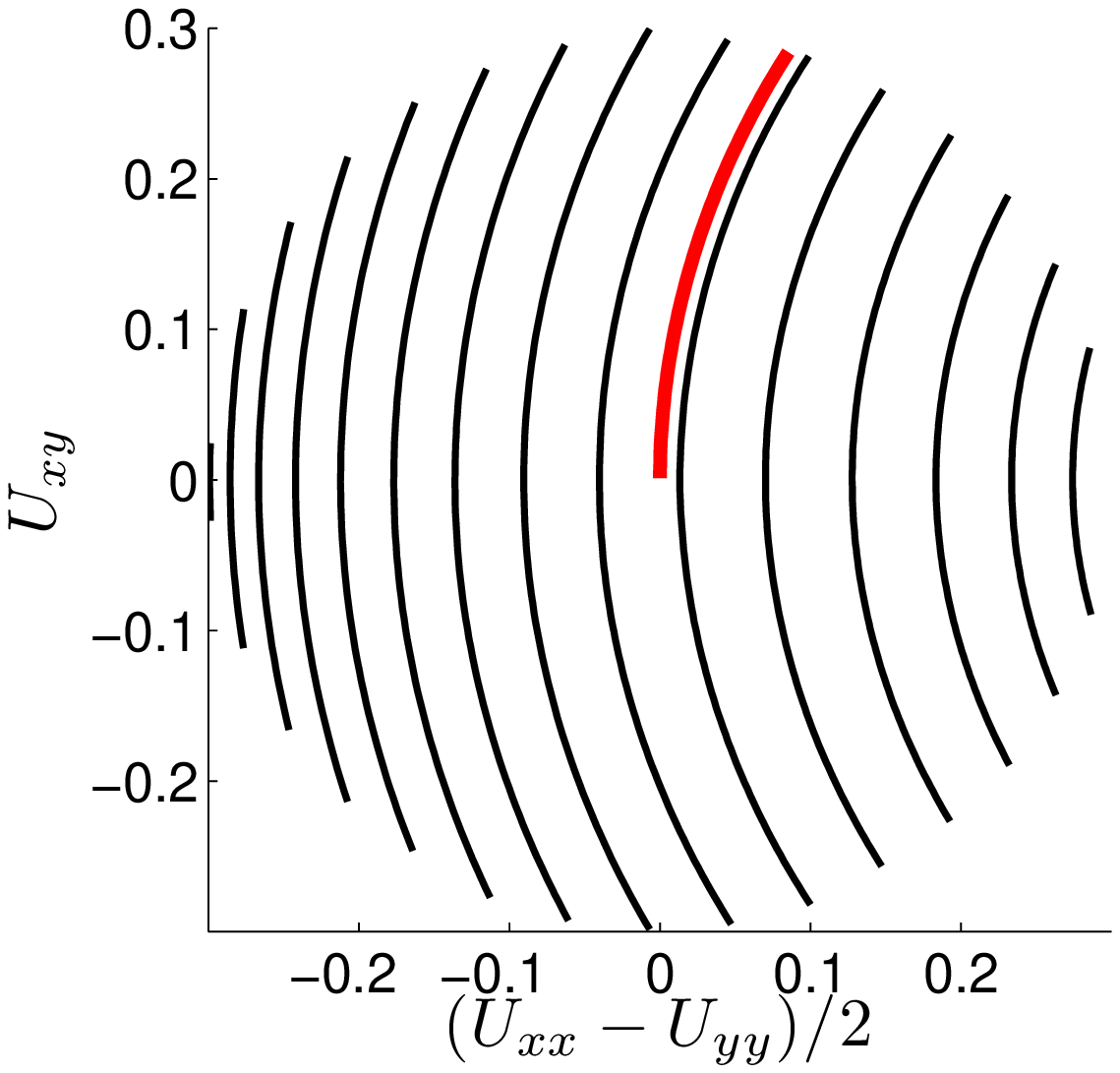}}
\put(0,0){\large{\textbf{a)}}}
\put(4.3,0){\large{\textbf{b)}}}
\end{picture}
\caption{Elastic model, represented
 in \emph{physical} (a) and \emph{component} (b)  spaces. Different initial states are taken:  $U_i = 0$ (center, red) and $U_i = 0.3$ (initial points scattered around the circle, black).  
Here $n \rightarrow +\infty$: all the trajectories evolve elastically.
For $\dot{\gamma} > 0$,
 $U_{xy}$ increases, that is, time evolve upwards on (b).
If $\dot{\gamma} < 0$,   these graphs  are unchanged, due to their symmetry with respect to the horizontal axis, showing that the purely elastic trajectories  are reversible.
 }
\label{Fig:ModelCurvesUyElast}
\end{figure}

As a first example of our representation, 
we consider here the pure elastic regime. This means that we allow the structure to deform elastically (stretching, contraction), but not to relax plastically (no rearrangements).
Our formalism allows us to describe pure elasticity by computing the elastic strain and its evolution when the material is deformed.
Our formalism extends to large strains, even those of order one (for strains much larger than one, without plasticity, the formalism of large amplitude strain  \cite{Labiausse2007} might be preferable). 
 Fig. \ref{Fig:ModelCurvesUyElast}ab shows trajectories for different initial elastic strains. 
 
 For an initially isotropic material, $U_i = 0$, we recover the classical results in the small strain limit: $U \simeq \gamma/2$ and $U_{xy} \simeq \gamma/2$. The Poynting relation  \cite{Labiausse2007} thus takes the form of a parabola; we even extend it to an initially anisotropic material, $U_i \neq 0$ (see Fig.  \ref{Fig:test} and \ref{Eq:ApproximationComplete}).
 For higher strains, $U_{xy}$ is less linear with respect to $\gamma$, due to the rotation of the elastic strain.

 In all cases, $U_{xy}$ increases monotonically. Note that  this is not the case for $(U_{xx}-U_{yy})/2$ nor for $U$.
 When the structure is aligned perpendicularly to the shearing direction ($\tensor{U}:\tensor{\nabla v}_{sym} < 0$), it contracts ($U$ decreases) under shear, until it aligns with the shear.
When the structure is aligned parallel to the shearing direction ($\tensor{U}:\tensor{\nabla v}_{sym} > 0$), it stretches ($U$ increases) under shear. 
Since $\tensor{U}$ is a tensor, it can continuously decrease, change direction and increase again without ever vanishing (as opposed to a scalar, which can change sign only when it is equal to zero).
For instance, a trajectory which starts with a direction opposed to that of shear has first a decreasing $U$  (contraction, with $U_{xy}$ negative and increasing), then an increasing $U$  
(stretching, with $U_{xy}$ positive and increasing), then a constant $U$ (yielding, with a rotation of $\tensor{U}$ towards the plastic limit).

\begin{figure}[!h]
\setlength{\unitlength}{1cm} 
\centering
\begin{picture}(7,5)(0.0,0.0)
\put(0,-0.2){\resizebox{7cm}{!}{\includegraphics{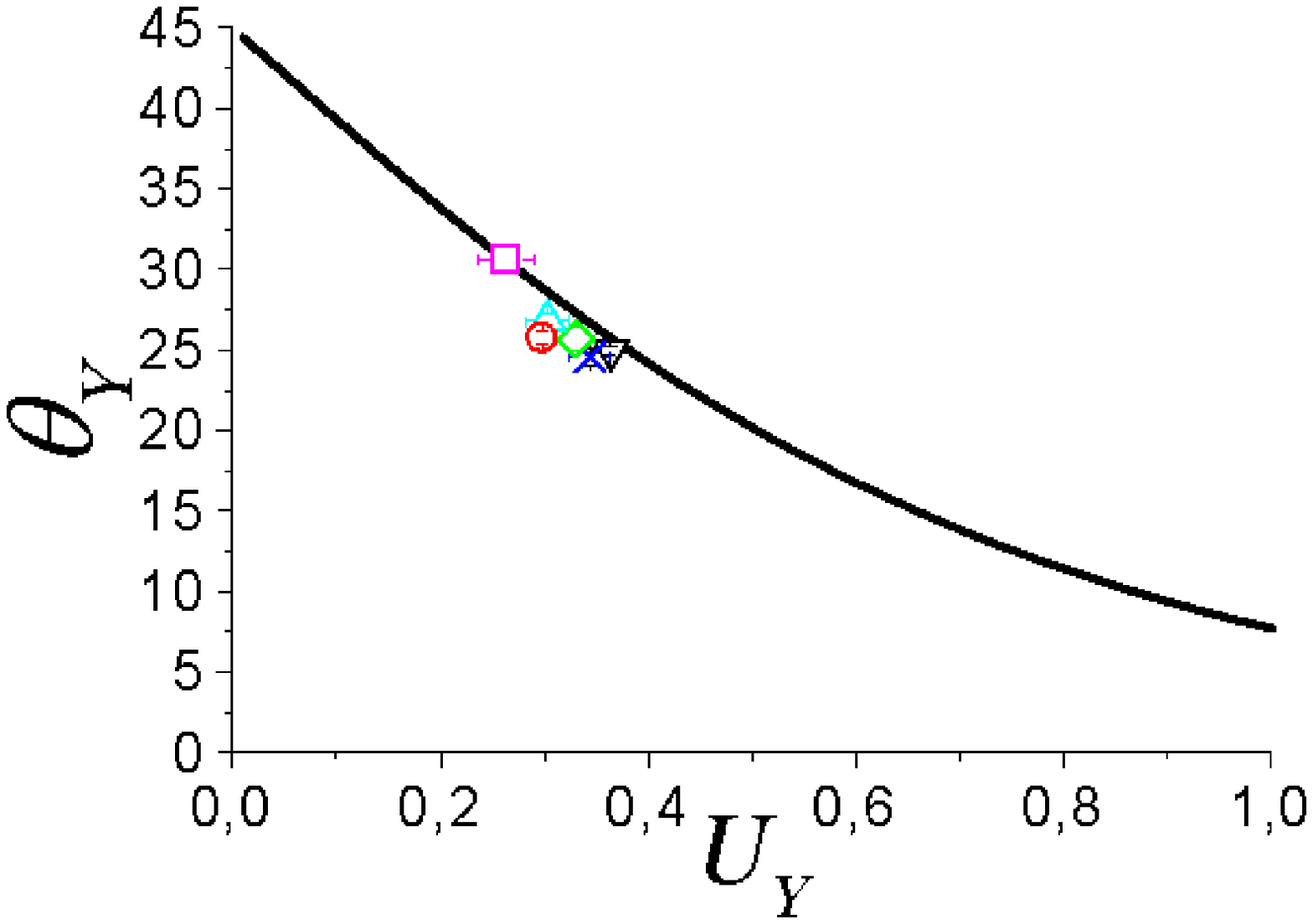}}}
\put(1.4,0.8){\resizebox{5.5cm}{!}{\includegraphics{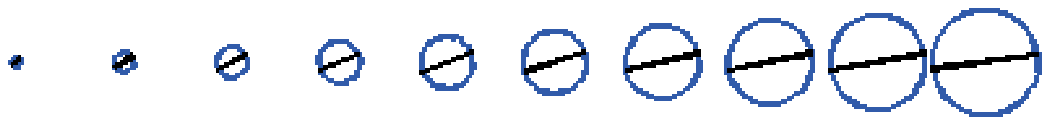}}}
\end{picture}
\caption{Plastic limit for $\dot{\gamma}>0$. Model of $\theta_Y$ versus $U_Y$ (solid line)
and corresponding representations of $\tensor{U}$ as circles with straight lines, indicating the direction of positive eigenvalue (inset), for several values of $U_Y$. 
Simulation points  (same symbols as in  Table  \ref{Tab:DefSimul}) are plotted for comparison.}
\label{Fig:PlaPlasticLimit}
\end{figure}

\subsubsection{Plastic Limit}
\label{par:modelPlasticLimit}

The yield strain is the amplitude of the strain when the material yields, that is, a scalar number. 
 The {\it plastic limit} is defined as the elastic strain tensor $\tensor{U}$ obtained after an infinitely long shearing ($\gamma \rightarrow +\infty$).  The amplitude of this tensor is that of the yield strain. Its direction is obtained by solving Eq. \ref{Eq:EvolutionEquation} 
 when its left-hand side  equals 0, $h$ equals 1, and ${\cal H}$ equals 1:
\begin{subeqnarray}
\label{Eq:PlaSolutionplastique}
\label{Eq:PlaSolutionplastique_1}
U & = & U_Y,\\
\label{Eq:PlaSolutionplastique_2}
\cos\theta &  = &  \frac{1}{\sqrt{1+e^{-4 U_Y}}},\\
\label{Eq:PlaSolutionplastique_3}
\sin\theta & = & \mathtt{sign}(\dot{\gamma})\frac{1}{\sqrt{e^{4 U_Y}+1}}
.
\end{subeqnarray}
This plastic limit is represented on Fig. \ref{Fig:PlaPlasticLimit}. 
It shows that the larger $U_Y$, the less aligned $\tensor{U}$ is with respect to $\tensor{\nabla v}_{sym}$. This tensorial effect is strong because $\theta_Y$ decreases quickly with $U_Y$. The scalar limit corresponds to $\theta \simeq 45^\circ$, as discussed in Sec. \ref{sec:ComparisonScalarTensorial}.

\subsubsection{Transient regime}
\label{sec:TransientRegim}

\begin{figure}[!h]
\setlength{\unitlength}{1cm} 
\centering
\begin{picture}(8,12)(0.0,0.0)
\put(0,8){\includegraphics[width=4cm]{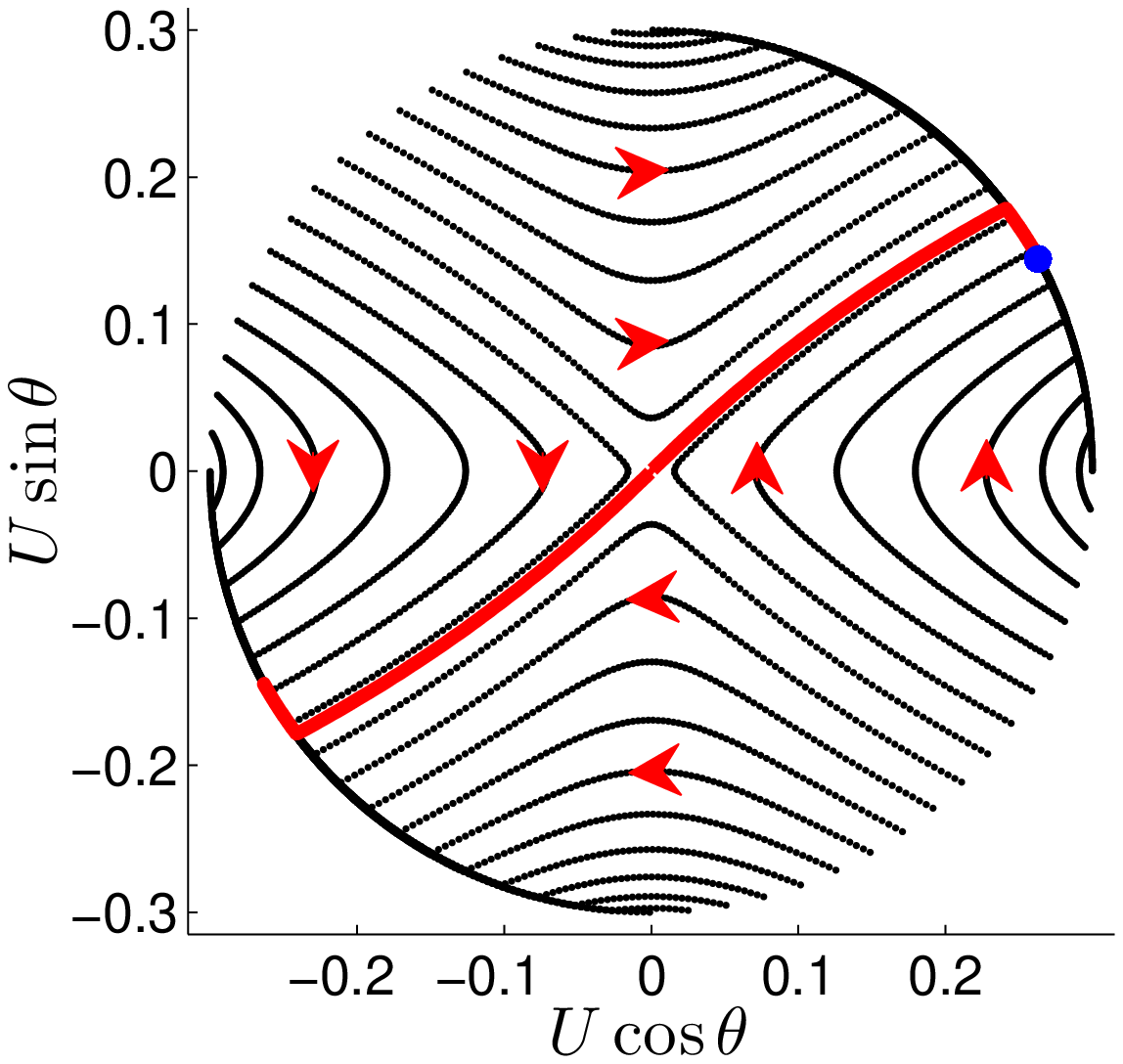}}
\put(4,8){\includegraphics[width=4cm]{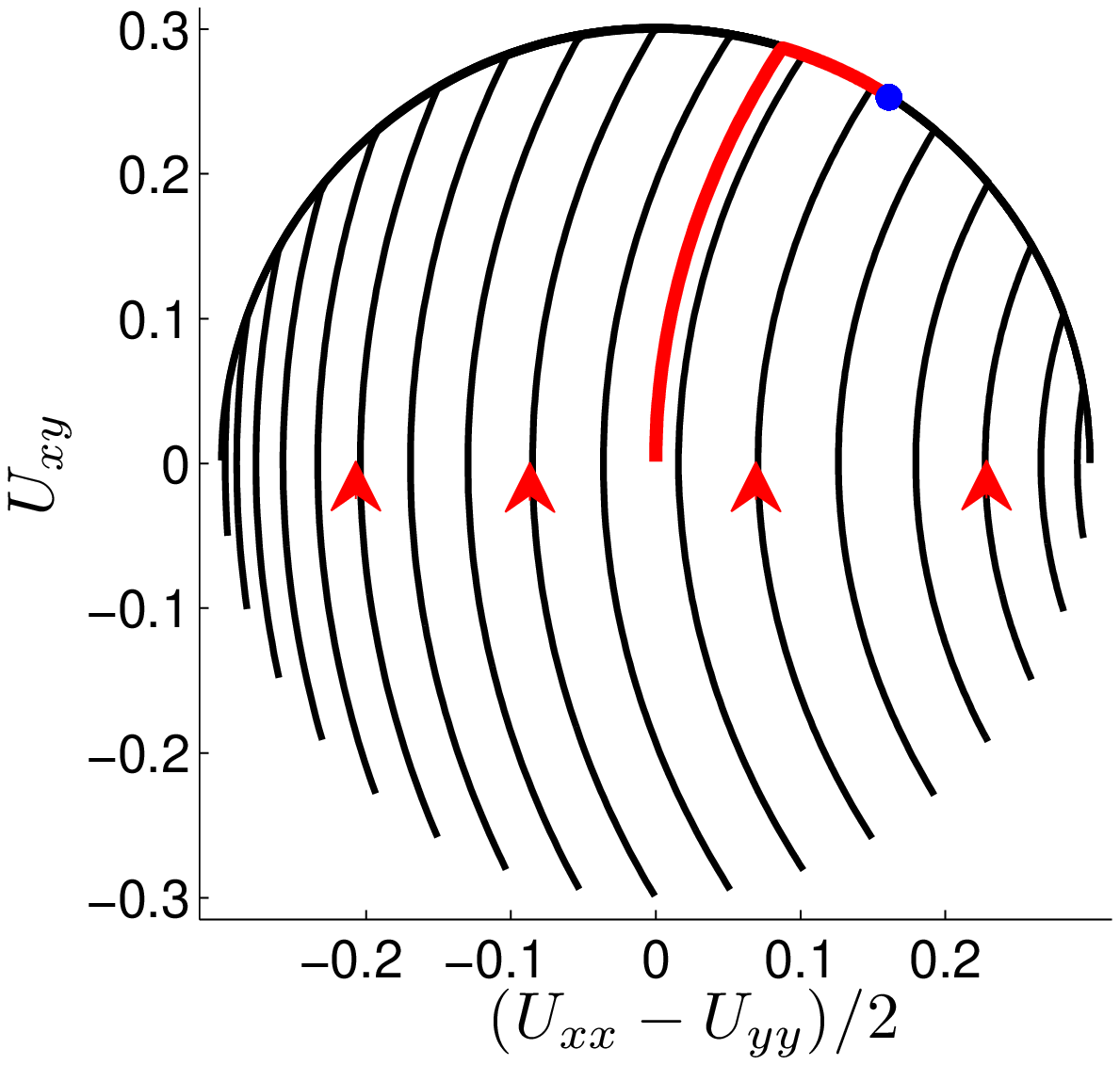}}
\put(0,4){\includegraphics[width=4cm]{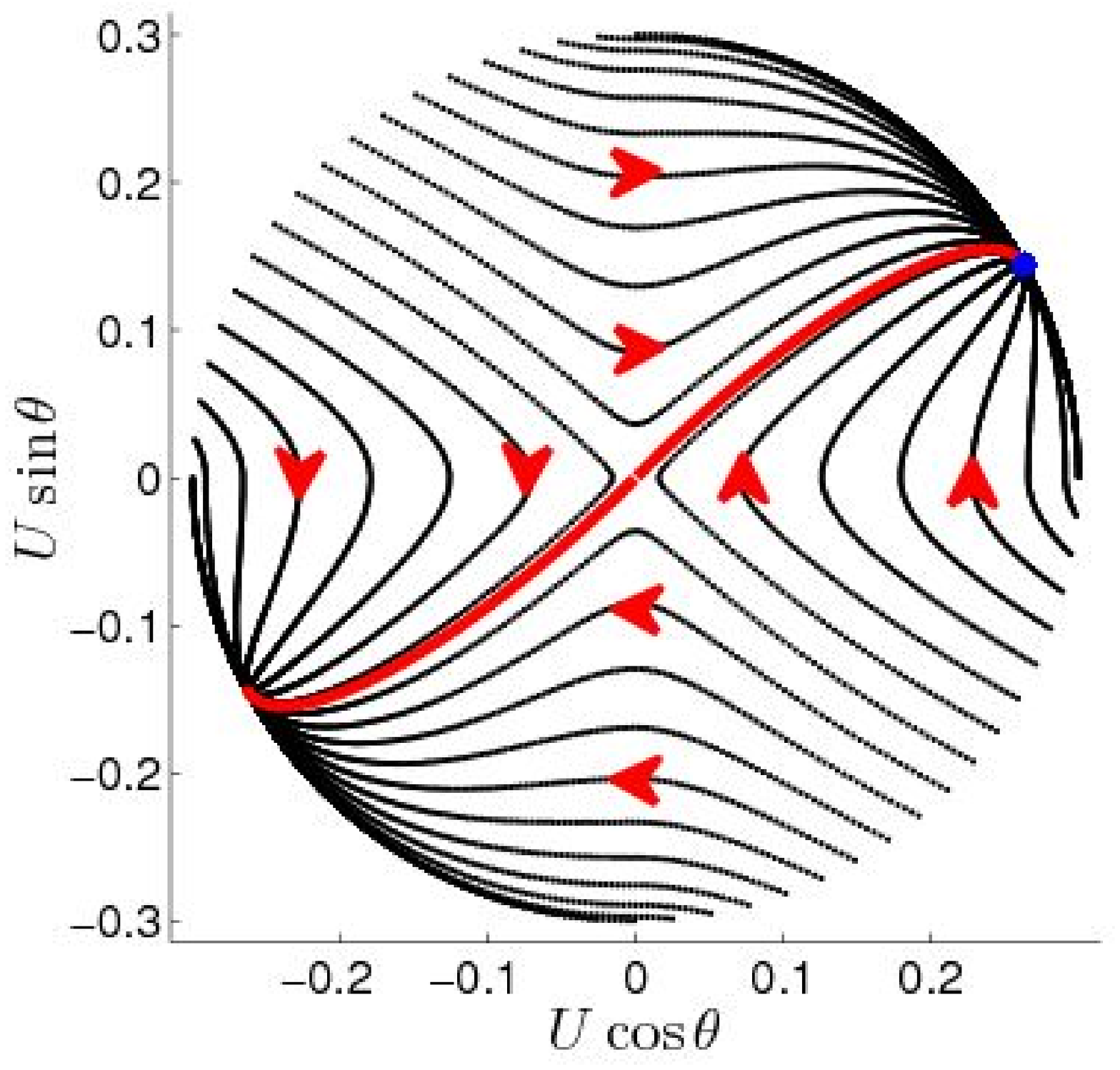}}
\put(4,4){\includegraphics[width=4cm]{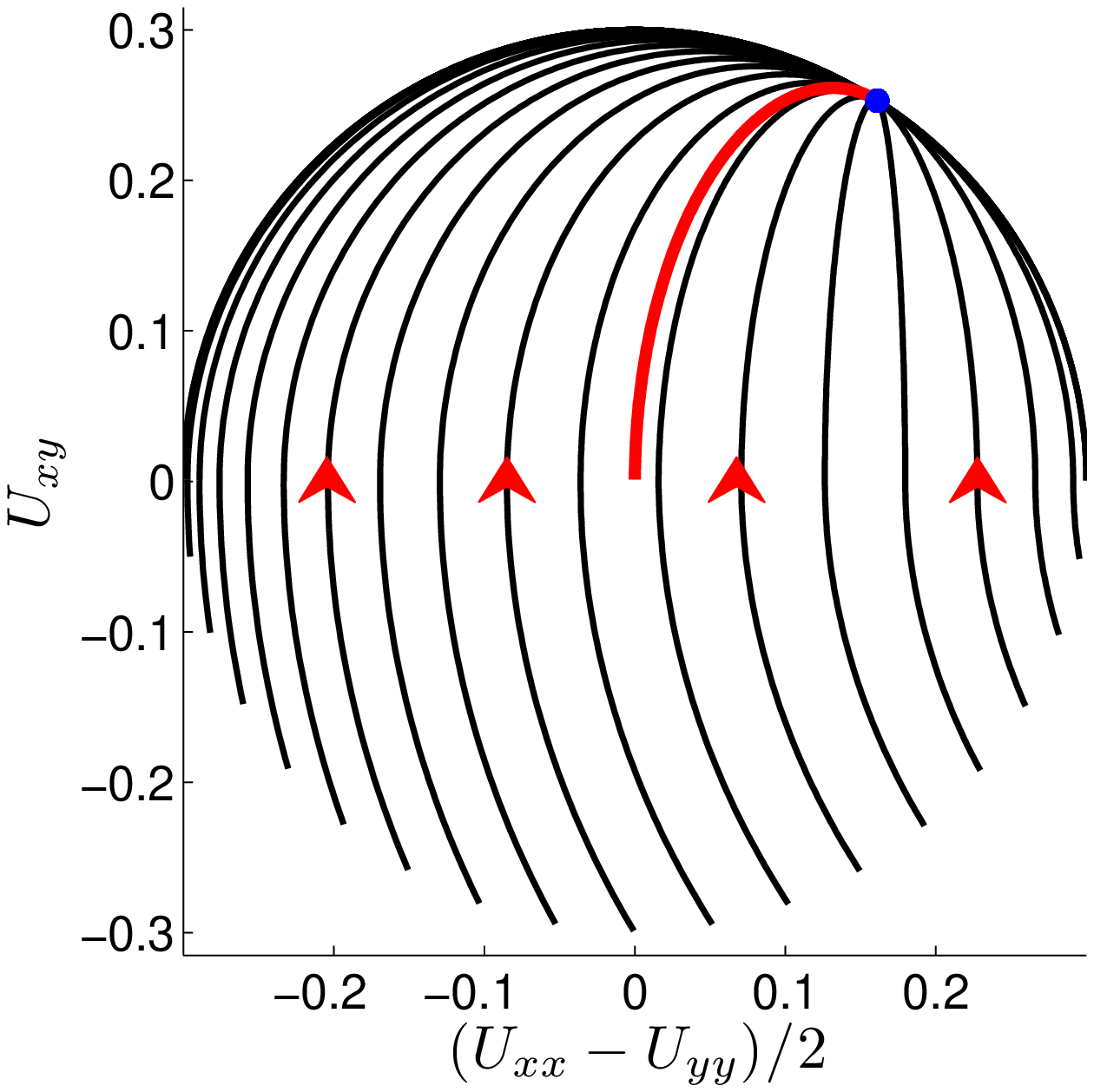}}
\put(0,0){\includegraphics[width=4cm]{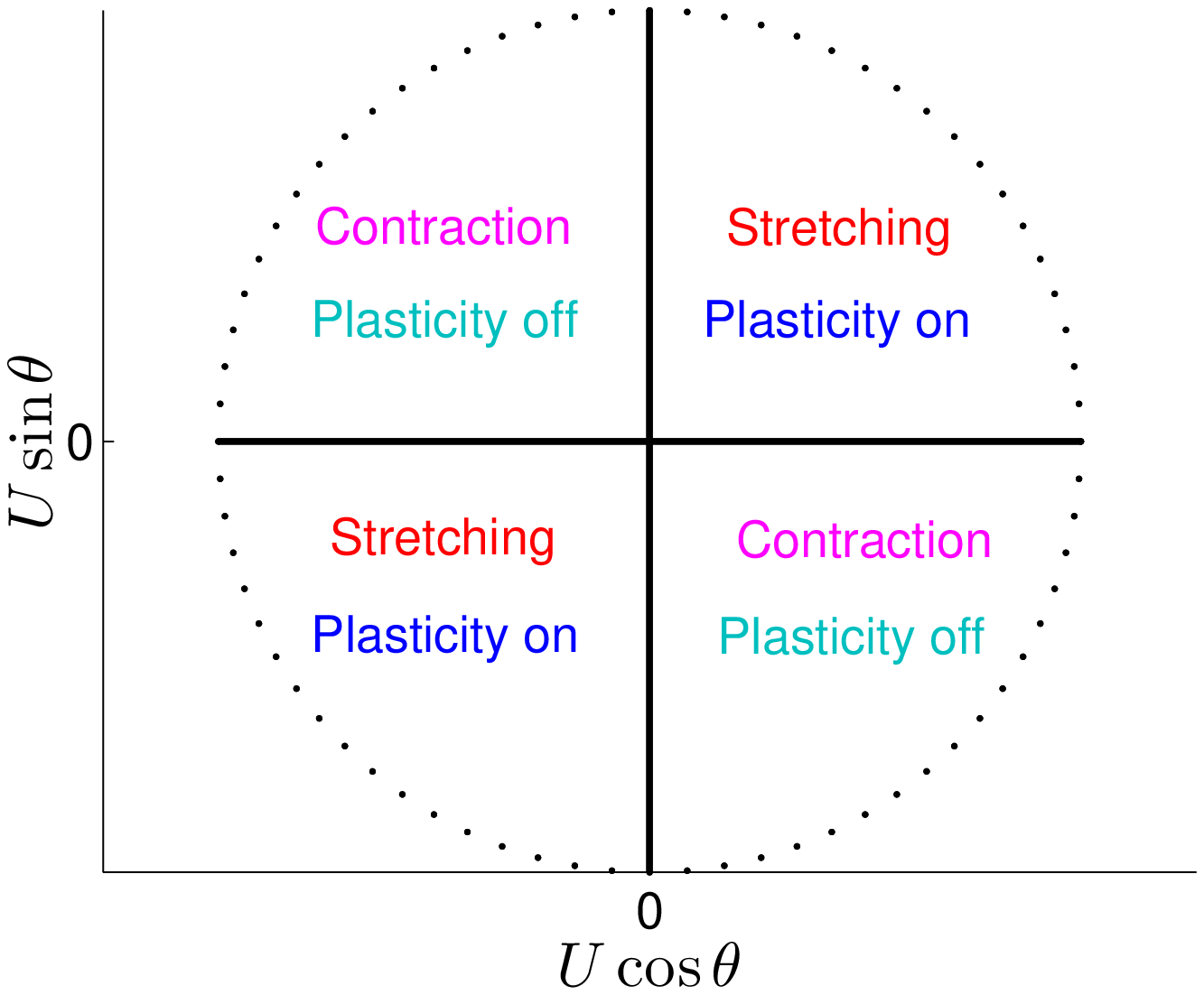}}
\put(4,0){\includegraphics[width=4cm]{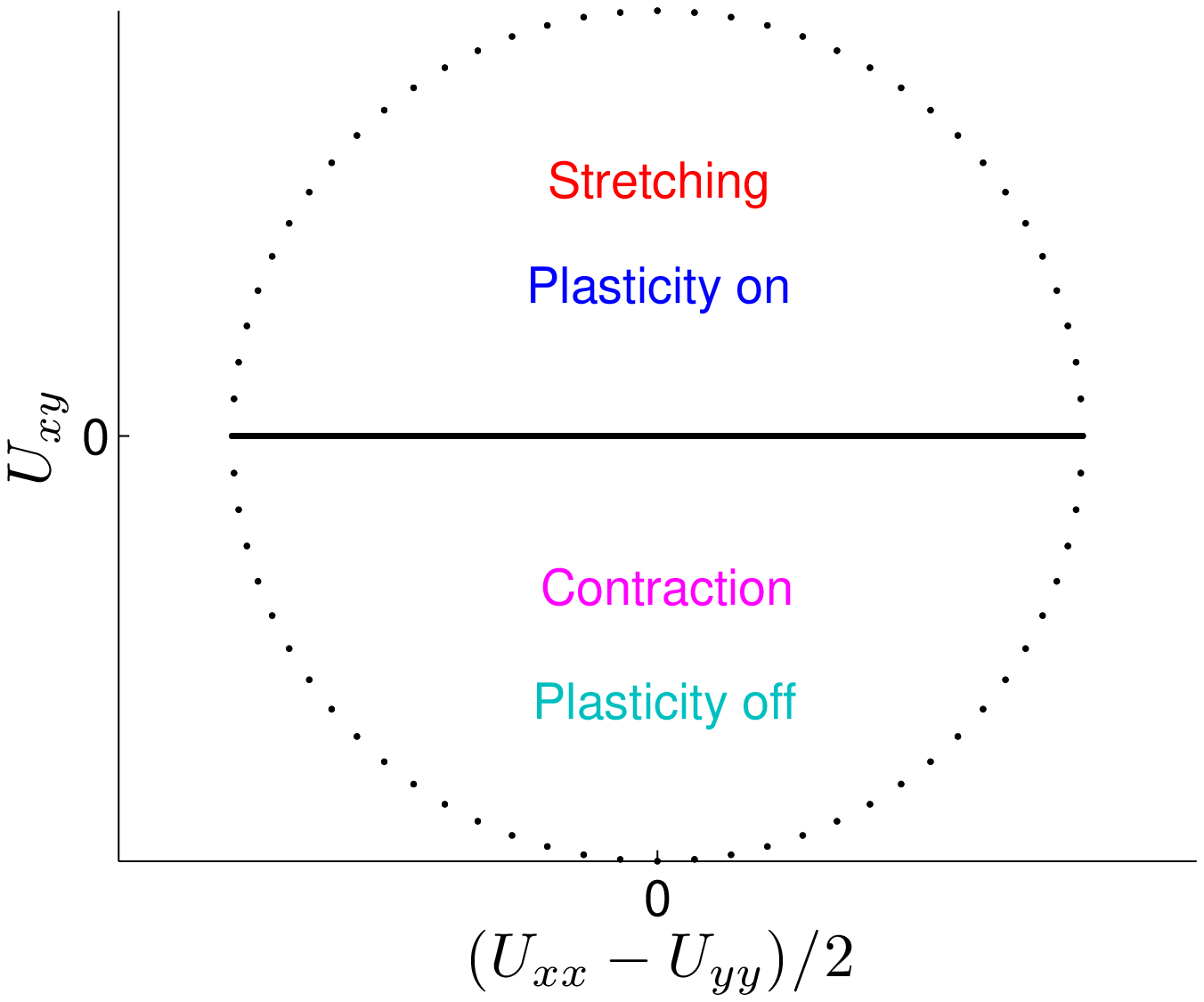}}
\put(0,8){\large{\textbf{a)}}}
\put(4,8){\large{\textbf{b)}}}
\put(0,4){\large{\textbf{c)}}}
\put(4,4){\large{\textbf{d)}}}
\put(0,0){\large{\textbf{e)}}}
\put(4,0){\large{\textbf{f)}}}
\end{picture}
\caption{Elasto-plastic model.
Representation of the model for $\dot{\gamma} > 0$ in \emph{physical} (a,c,e) and \emph{component} (b,d,f)  spaces. 
  If $\dot{\gamma} < 0$, the vertical axis of all these graphs should be inverted.
(a-b) For $n \rightarrow +\infty$, all the trajectories evolve elastically ($U_{xy}$ increases, that is, time evolves in the direction of the arrows) up to the yield strain, here taken as $U_Y=0.3$, then evolve plastically ($U$ constant) towards
 the plastic limit (blue point) which corresponds to a specific angle, see Eq. 
 \ref{Eq:PlaSolutionplastique}. If the trajectory reaches the yield strain on the left of the plastic limit, $U_{xy}$ passes through a maximum (overshoot). 
 (c-d) For $n = 2$, plasticity appears more progressively, thus smoothening the transition between elastic and plastic regimes, and decreasing (or even suppressing) the overshoot. 
  (e-f) Summary of the mechanical behavior: ``stretching" and ``contraction" according to the direction of $\tensor{U}$ with respect to shear. Here ``plasticity on" or ``plasticity off" refers to the Heaviside function in the last term of Eq. \ref{Eq:EvolutionEquation}, when the plasticity is progressive ($n$ finite); when $n$ increases the ``plasticity on" zone narrows, and for $n$ infinite it is reduced to the limit circle.
}
\label{Fig:ModelCurvesUyPlast}
\end{figure}

We now consider the shearing of an initially anisotropic elastic strain, $\tensor{U}_i\neq \tensor{0}$. In \emph{physical} or \emph{component space},  the state of the material is initially situated on an elastic trajectory and must arrive at the plastic limit point (Fig. \ref{Fig:ModelCurvesUyPlast}a-d). At this limit point, an increase of elastic strain is immediately transformed into plastic strain. Plasticity may occur only if $\tensor{U}:\tensor{\nabla v}_{sym} > 0$.

Graphically, both in \emph{physical space} and in \emph{component
space} (Fig. \ref{Fig:ModelCurvesUyPlast}),  the plastic limit is
represented by the point where a trajectory reaches perpendicularly the
circle at $U = U_Y$ (the elastic strain increases along the
tangent of the trajectory, the plastic strain relaxes towards the centre of
the circle: to balance each other, they must be parallel).

The shape of the yield function $h$ then  determines how the material reaches the plastic limit.  For simplicity, we take $h$ as a power law function: $h = (U/U_Y)^n$ (Fig. \ref{Fig:Pla-h-n}a). Two examples of the resulting behaviour are plotted on  Fig. \ref{Fig:ModelCurvesUyPlast}.
The limit $n \rightarrow \infty$ is intuitive: the material follows the elastic trajectory up to $U = U_Y$; then $U$ is fixed and the plastic limit is reached by describing an arc of a circle
in  \emph{physical} and \emph{component spaces} (Fig. \ref{Fig:ModelCurvesUyPlast}ab). 
For other cases (finite $n$) plasticity occurs earlier (Fig. \ref{Fig:ModelCurvesUyPlast}ef) and trajectories converge to the plastic limit (Fig. \ref{Fig:ModelCurvesUyPlast}cd). 
 
The behaviour changes qualitatively if the sign of $\dot{\gamma}$ is abruptly reversed. Unlike the elastic term, the plasticity term is irreversible due to  the Heaviside function $\cal H$ in 
Eq. \ref{Eq:EvolutionEquation}.
This leads to an inversion of the plastic domain in \emph{physical} and \emph{component spaces} (Fig. \ref{Fig:ModelCurvesUyPlast}ef) and to a new plastic limit position ($\theta_Y \rightarrow -\theta_Y$, Eq. \ref{Eq:PlaSolutionplastique}). This hysteretic effect is shown on  
Fig. \ref{Fig:SimulationRepresentation} 
by reversing $\dot{\gamma}$ once the plastic limit is reached. For high $n$,  the new plastic limit is quickly reached, since the two plastic limits are on the same elastic trajectory. 

If we perform alternate sign changes of $\dot{\gamma}$, we observe that the material is stuck in a limit trajectory
(Fig. \ref{Fig:Pla-h-n}b).
This trajectory is almost insensitive to $h$ and therefore close to the elastic trajectory joining the two plastic limits. This has an important consequence: once in the plastic regime, the elastic strain (and thus the stress) can not be totally relaxed if we only reverse the shearing direction. This is examined in more detail below (Fig. \ref{Fig:RelaxationCycles}).

\begin{figure}[!h]
a)
 \includegraphics[width=7cm]{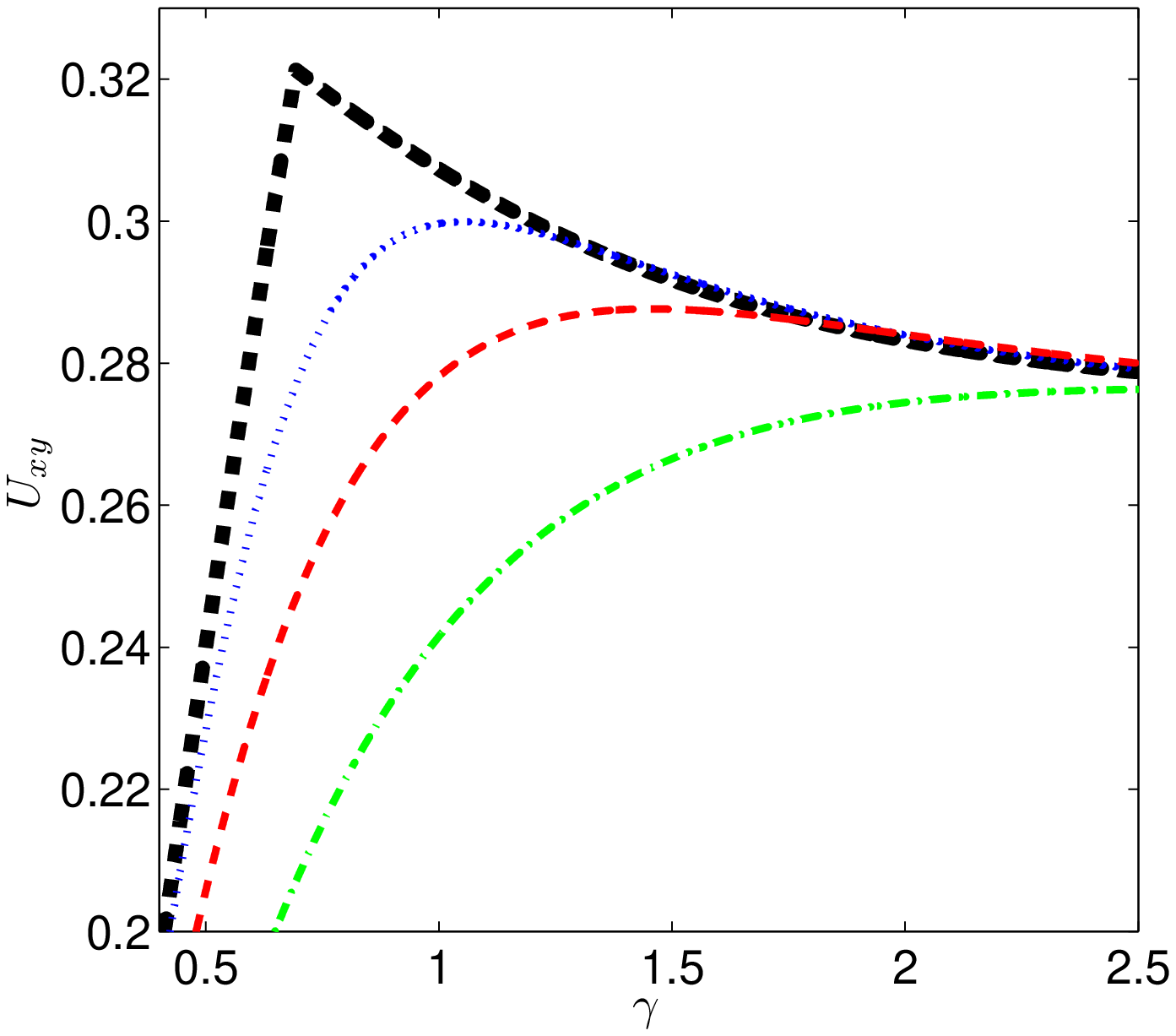}\\
b)
 \includegraphics[width=7cm]{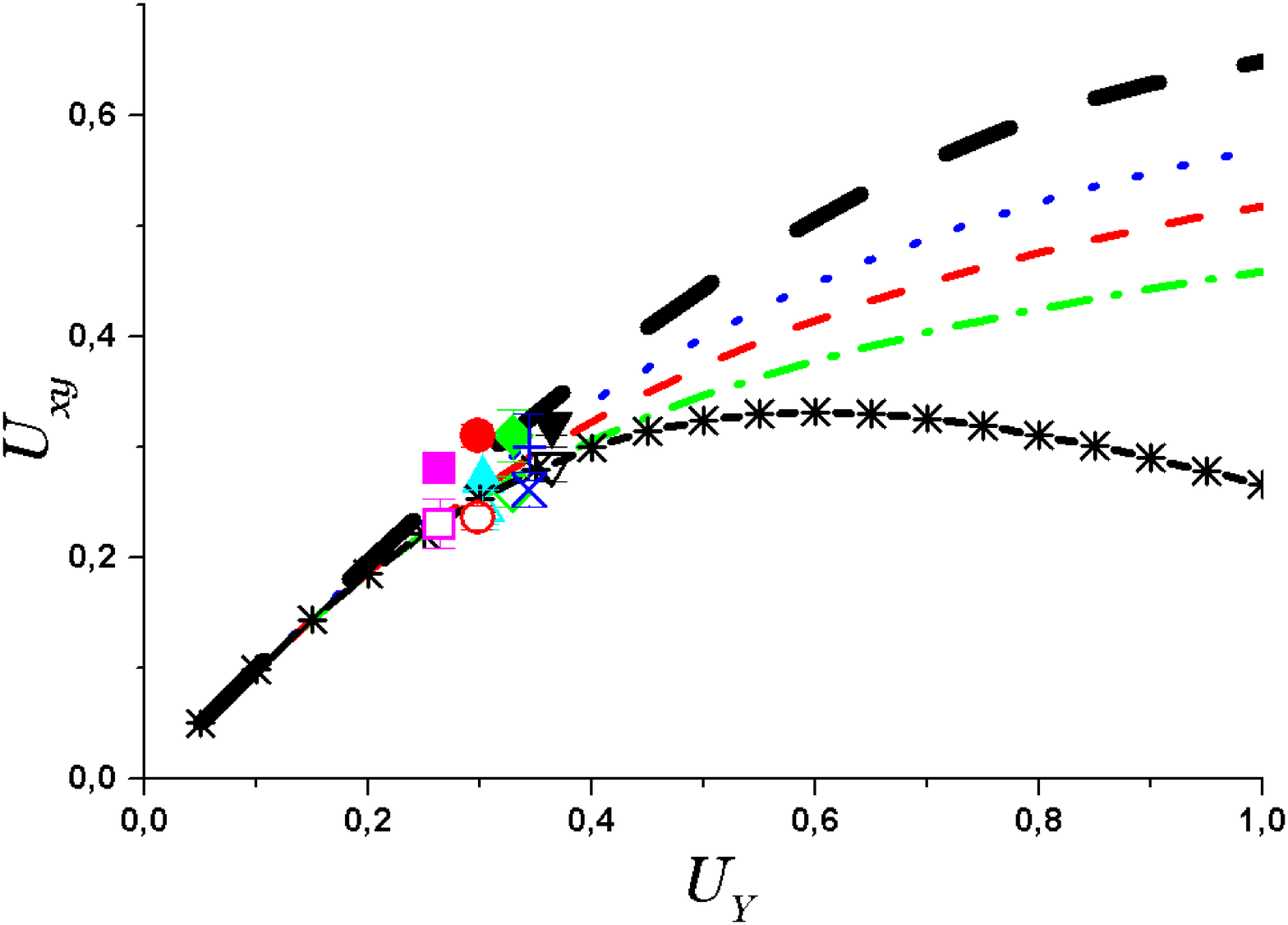}\\
c)
 \includegraphics[width=7cm]{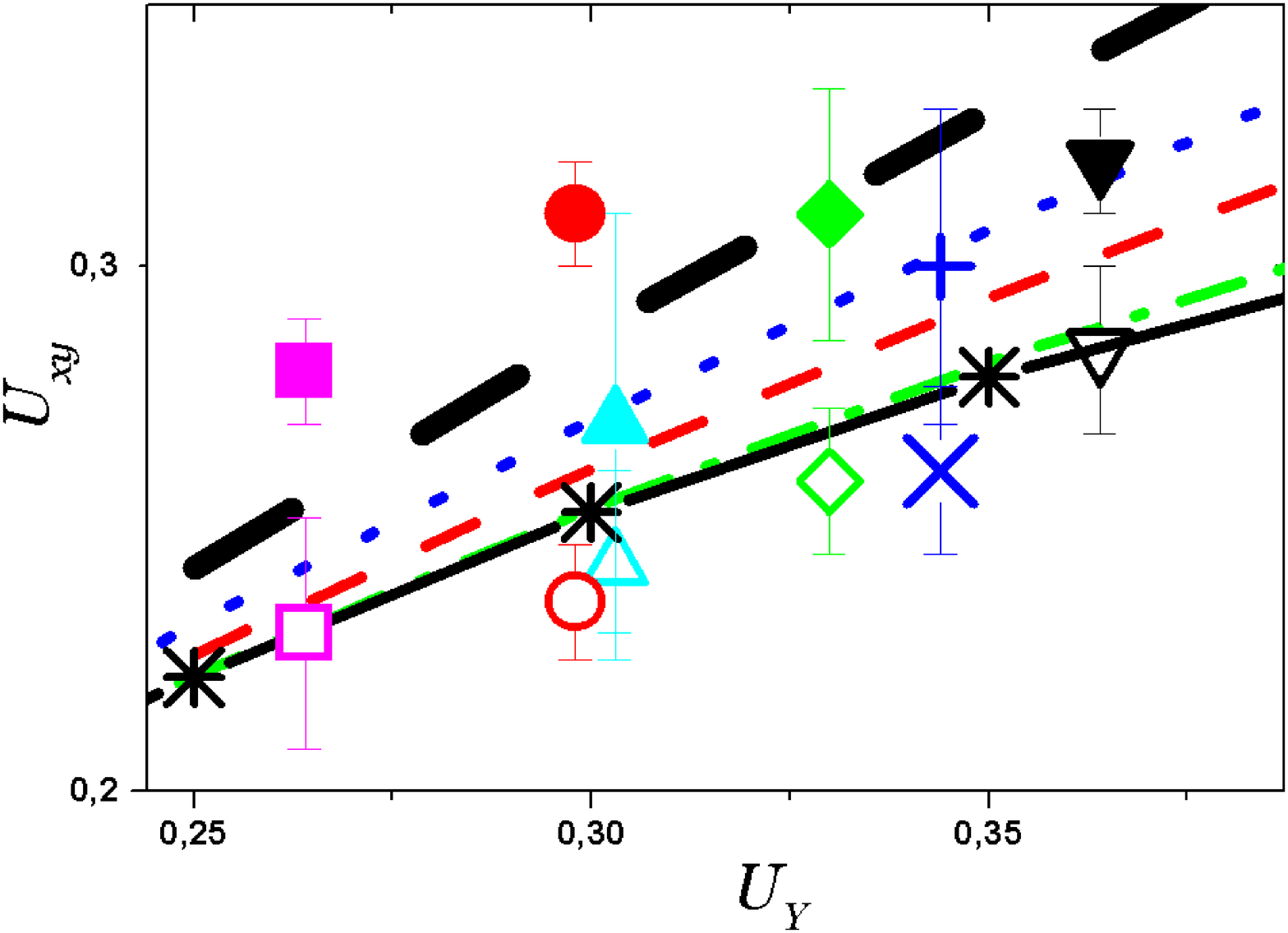}\\
\caption{
Overshoot, defined as the difference between the maximum of  the $U_{xy}$ versus $\gamma$ curve, and the value of the plateau which follows this maximum. 
Calculations are performed in the case of an initially isotropic structure ($U_i = 0$).
(a)  Zoom over the maxima of the curves of 
 Fig. \ref{Fig:SimulationRepresentation}a.
(b) Maxima for different $h$ functions (same legend as Fig. \ref{Fig:Pla-h-n}), and plateau value (starred line, which is the same for all $h$ functions), plotted {\it vs} $U_Y$. 
 Simulation points 
 are plotted for comparison, with the same symbols as in Table \ref{Tab:DefSimul}; closed symbols and  {\color{blue}$+$}  correspond to the maximal value (averaged over a few successive points) during the first shear step; open symbols and {\color{blue}$\times$} correspond to the averaged value along the plateau of the last shear step; $U_Y$ is measured as the plateau value of $U$.
 (c) Zoom of (b).
}
\label{Fig:PlaOvershootUs}
\end{figure}

\subsubsection{Overshoot}
\label{sec:Overshoot}
   
As observed for $n \rightarrow +\infty$ (Fig. \ref{Fig:ModelCurvesUyPlast}b),
the overshoot is due to the transition from an elastic trajectory to the plastic limit. 
The structure itself has no overshoot: $U$ increases monotonically.
The overshoot appears in the tangential strain $U_{xy}$: it is   a purely tensorial effect due to a rotation of the structure. 
In fact, $U_{xy}$ increases in all elastic trajectories.
Upon reaching $U=U_Y$ there is a sudden transition to the plastic regime. For trajectories to the  left of the plastic limit (Fig. \ref{Fig:ModelCurvesUyPlast}b), 
we see that $U_{xy}$ decreases towards the plastic limit. The overshoot corresponds to the difference between the maximum value of $U_{xy}$
(where the trajectory meets the circle) and the plateau value (plastic limit). 

>From Fig. \ref{Fig:ModelCurvesUyPlast}b, 
we observe a tiny overshoot for the normal stress difference  if
   the trajectories reach the $U=U_Y$ circle to the right of the plastic limit. In that case, 
   the trajectories move towards the right in the elastic regime, then towards the left in the plastic regime.
Such trajectories correspond to   the right of Fig. \ref{Fig:ModelCurvesUyPlast}b, that is, a structure with a large trapped normal stress difference 
$U_{xx}-U_{yy}$.

For smaller $n$, the elastic-plastic transition is smoother and the overshoot is reduced.
The overshoot amplitude for different $h$ and $U_Y$ is plotted  on Fig. \ref{Fig:PlaOvershootUs}, 
for the case of an initially isotropic structure ($U_i = 0$).
The overshoot increases with $U_Y$ because the plastic limit moves away from the initial elastic trajectory. 

\section{Comparison between simulations and the model}
\label{sec:ComparisonSimulationModel}

\subsection{Plastic limit}
\label{sec:ResultPlasticLimit}

The model  (section \ref{sec:TransientRegim}) predicts that after a few cycles a limit trajectory  is reached. This trajectory is also displayed in simulations by Kabla and Debregeas \cite{KablaPreprintSimul}.
The plastic limit can thus be evaluated in simulations:  $U_Y$ is estimated by averaging $U$ on the last plateau 
 (using any of Fig. \ref{Fig:SimulationRepresentation}a-d);
similarly, $\theta_Y$ is estimated by averaging $\theta$ on the last plateau of Fig. \ref{Fig:SimulationRepresentation}b or d.
On Fig. \ref{Fig:PlaPlasticLimit}, results from simulations are compared with the model. Agreement is good, and the model captures tensorial effects, especially because the measured  $\theta_Y$ deviate much from the $45^\circ$ scalar limit.
  
Taking larger $l_c$ in the simulations favours neighbour swappings (``T1s"), thus it corresponds to an increased effective liquid fraction. As expected   \cite{{Saint-Jalmes1999},Raufaste2007}, we see a decrease in $U_Y$. However, since the effective liquid fraction we are simulating remains in a very dry range ($<4 \; 10^{-4}$), it does not influence much $U_Y$, which thus varies over a narrow range ($0.26-0.37$). 

Reaching higher $U_Y$ is possible with other materials, but not with disordered 2D  foams. Reaching lower $U_Y$ is possible (and usual) in experiments on disordered wet foams, but not in the present simulations where the algorithm would require adaptation at high $l_c$.
 
\subsection{Yield strain and yield function }
\label{sec:EvolutionOfTheStructure}
 
Simulation results fluctuate, due to the limited number of bubbles (discrete description), while model curves are smooth, corresponding to the limit of a large number of bubbles (continuous material description). 
There is a qualitative agreement, which is good enough to deduce  $U_Y$ and $h$ approximately. 

For instance, Fig. \ref{Fig:SimulationRepresentation}  compares a simulation with models using various $h$ functions. 
We observe that $n \approx 2$ 
 (red dashes on Fig.  \ref{Fig:SimulationRepresentation})
describes well the simulation during the first positive shearing step, and $n \approx 4$ 
 (blue dotted lines on Fig.  \ref{Fig:SimulationRepresentation})
during the second one.
Similarly, $U_Y$ is deduced from the plateau value of $U$ (Figs.   \ref{Fig:PlaPlasticLimit} and \ref{Fig:PlaOvershootUs}).

In practice, in a first approximation, it is enough to consider  $U_Y$ and $h$ as constant. 
Their variations are small and thus have a small effect on the foam rheology. 
However, these variations do exist. 

For instance, in this example of Fig. \ref{Fig:SimulationRepresentation},  $n$ (and thus $h$)  evolves throughout the simulation, revealing that the structure 
evolves too;  $h$ seems to be sensitive to the (topological) disorder of the foam  \cite{{QuillietPreprint}}. Here $U_Y$ is constant, but there are other cases  (data not shown, see \cite{RaufasteThese}) where, due to the decrease of the topological disorder during the shearing, $U_Y$ decreases.

More generally, a real foam is constantly evolving under the effect of drainage, coarsening
 \cite{Hohler2005}, or
shuffling \cite{{QuillietPreprint}}. 
These effects should probably have to be considered in future models, which would try to predict $U_Y$ and $h$, based on the average and fluctuations of the structure, respectively.

\subsection{Overshoot }
\label{sec:DiscOvershoot}

We can now identify two distinct physical mechanisms which can cause a stress overshoot in shear experiments of  elasto-plastic materials.

The first one is an {\it orientation} effect, suggested in Sec. \ref{sec:Overshoot}. 
$U$ increases monotonically, but if $U_Y$ is large enough then the rotation of $\tensor{U}$ under shear implies that the tangential shear strain $U_{xy}$ passes through a maximum. 
This purely tensorial effect is absent from scalar models. 
Under certain additional conditions on the initial elastic strain $U_i$, which are also described correctly only when taking into account the tensorial aspects, the normal strain difference $U_{xx}-U_{yy}$ too passes through a maximum.

Fig. \ref{Fig:PlaOvershootUs} shows a comparison between the model and the simulations. Given that in the range of simulated  $U_Y$ the overshoot is tiny and difficult to extract from the fluctuations, 
the agreement is surprisingly good. In most foam experiments, where  $U_Y$ is even lower, 
this effect should be too small to be measurable.

The second one is  outside of the scope of the present paper. 
It is due to an evolution of the {\it structure} itself during the first shear step (see section \ref{sec:EvolutionOfTheStructure}).
This might be invoked to explain the larger overshoot of the data corresponding to the confined simulations (red and pink), as well as most experimental observations (such as that of ref. \cite{Khan1988}).

\section{Practical applications}
\label{sec:Practical}

\subsection{Comparison between scalar and tensorial representations}
\label{sec:ComparisonScalarTensorial}
    
As long as the applied shear keeps a constant direction, and the elastic strain remains much smaller than 1, 
its eigenvectors correspond to that  of $\tensor{\nabla v}_{sym}$. That is, they are at $45^\circ$  to the direction of shear. 
This is called the {\it scalar approximation},
 and it considerably simplifies the  study of the mechanical behaviour. 
In that case, 
a single (scalar) number is enough to fully describe   the elastic strain.

This scalar number can equally well be chosen as the amplitude $U$, or the eigenvalue $U_1$, or the tangential shear strain $U_{xy}$, among others. To switch from one choice to the other requires
care regarding the prefactors  \cite{Cartes}: this is often a source of confusion in the literature, especially regarding the definition and value of the yield strain.
The link between the simplified (scalar) and complete (tensorial) equations is detailed in Appendix \ref{sec:AnnScalarLimit}, using $2U_{xy}$ as a scalar.

If $U_Y \ll 1$, which is the case for wet foams and emulsions, then $U$ remains always  much smaller than 1, and $\theta_Y \simeq 45^\circ$, so that the scalar approximation holds,  see Fig. 19 in  ref.  \cite{Cartes} 
(except if the direction of the shear changes, in 2D or in 3D).
 In that limit tensorial effects such as normal differences or stress overshoot are negligible.

  Quantitatively, $U_{xy}$ is linked to $\sin(2 \theta)$ (Eq. \ref{Eq:UPrincipalComponent_1}).
 This implies that a difference of 10\% between the scalar and tensorial equations is reached when $\sin(2 \theta_Y) = 0.9$, which corresponds to $U_Y = 0.23$ (Eq. \ref{Eq:PlaSolutionplastique}). 
 Very dry foams, such as those  simulated here, are slightly above this limit: a tensorial model is therefore useful. 

\begin{figure}[!h]
a)
 \includegraphics[width=7cm]{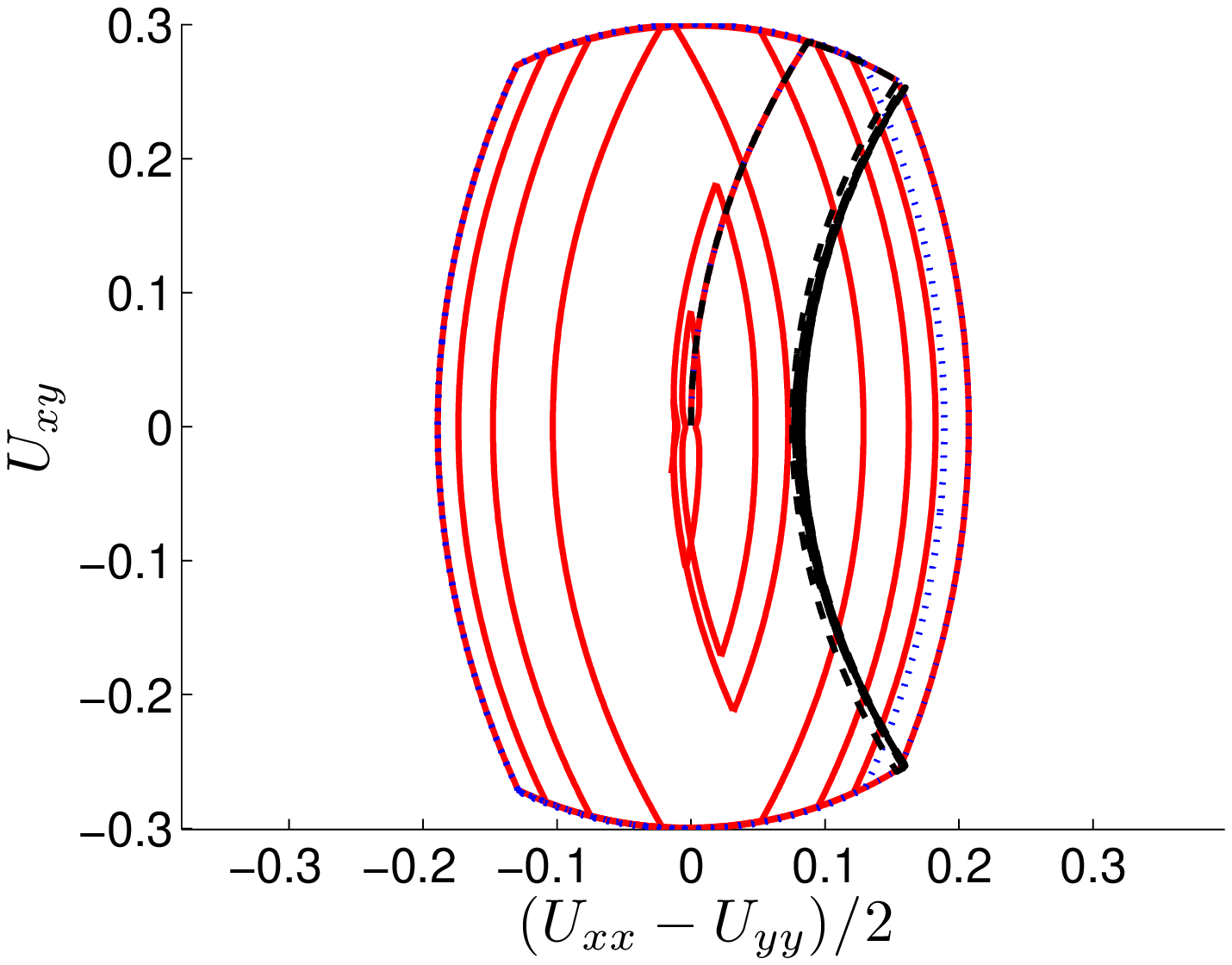}\\
b)
 \includegraphics[width=7cm]{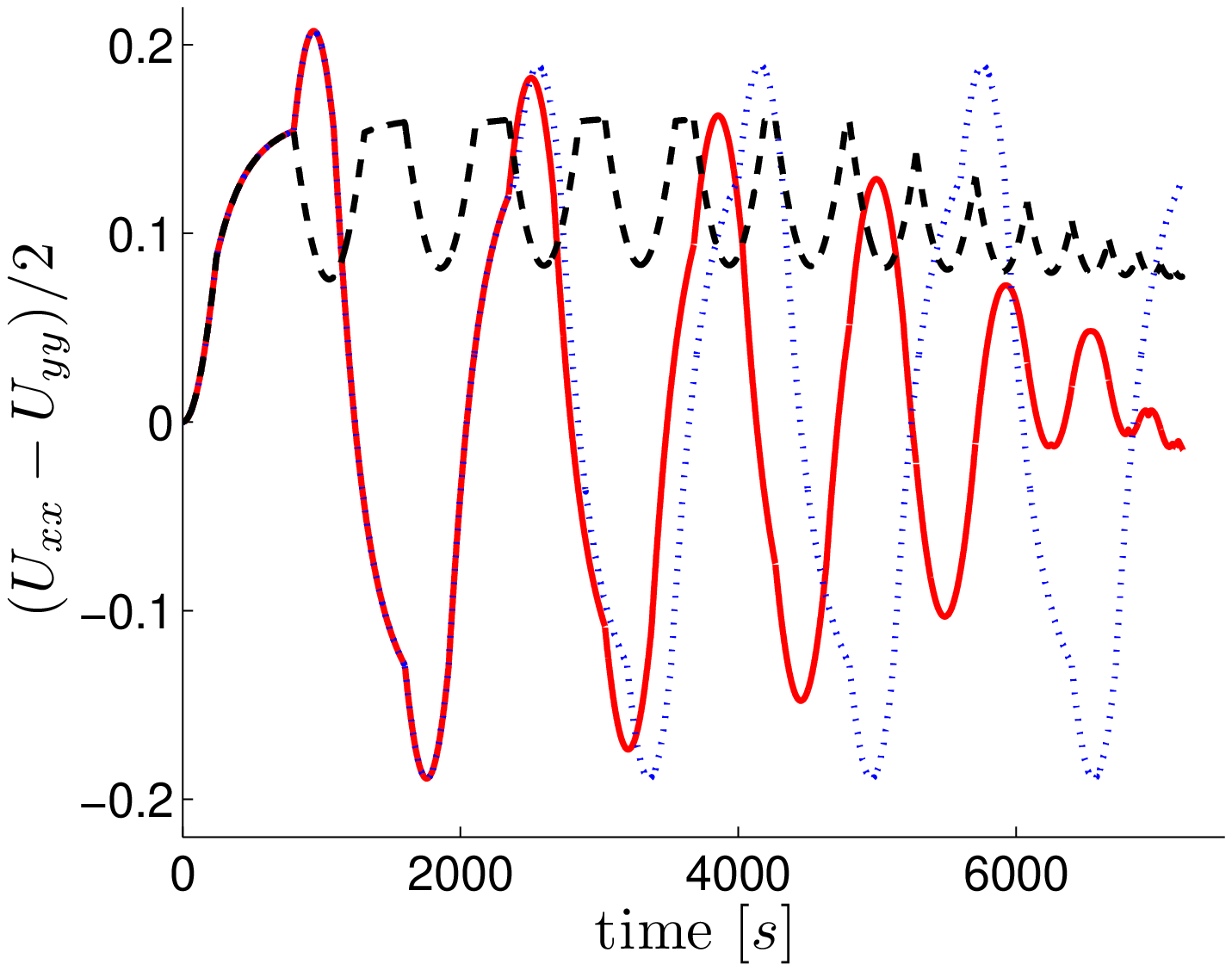}\\
\caption{Shearing cycles to remove trapped stresses. Here $U_i = 0$ for the first step and $U_Y = 0.3$, $n \rightarrow +\infty$. Between two steps:  the direction of shearing is turned $90^\circ$ clockwise (blue dots);  the amplitude is decreased from $\gamma =2$ to 0 in 15 steps (black dashes); simultaneously,  the direction of shear is turned and its amplitude is decreased (red solid line).
a) Component space. b) $(U_{xx} - U_{yy})/2$ versus time ($\left|\dot{\gamma}\right| = 2.5 \times 10^{-3}$s$^{-1}$).}
\label{Fig:RelaxationCycles}
\end{figure}

\subsection{Trapped strains and stresses}
\label{sec:TrappedStrainsStresses}

A dry foam is a material with sufficiently high $U_Y$ that normal stresses may exist even when the material is at rest \cite{Larson1999,Kraynik2003,Janiaud2005}.  To relax such residual (or ``trapped") stresses, 
we should   first shear the foam enough to reach the plastic stage, so that  plastic rearrangements anneal the disorder.
We then must perform cycles of shear.

If the direction of shear is kept constant, and the shear simply reversed, the foam asymptotically reaches a limit trajectory, and the stress is not relaxed. 
Decreasing the amplitude of the shear cycle does not enable to leave this limit trajectory (black dashes in Fig. \ref{Fig:RelaxationCycles}ab).
Kraynik \emph{et al.}   simulated dry 3D foam and applied shearing cycles (actually uniaxial contractions) of amplitude $\approx 0.2$  in different directions, rotated by $90^\circ$; this procedure decreases the trapped stress by  a factor of around 2, which does not improve with more cycles  (Fig. 7 of \cite{Kraynik2003}).

Here we propose a reproducible procedure based on section  \ref{sec:TransientRegim}, which couples shearing cycles in different directions and decreasing amplitudes, as follows:
\begin{itemize}
\item The amplitude $\gamma_i$ of the first step  is large enough to completely reach the plastic stage: $\gamma_i \gg 2 U_Y$.
\item At each step, the shearing direction is rotated by $90^\circ$
and the shearing amplitude is decreased.
\item The decrease in amplitude between successive steps is smaller than $2U_Y/5$, ensuring there are at least 5 steps between $2 U_Y$ and 0
({\it i.e.} the total number of steps is at least $5 \gamma_i / 2 U_Y$).
\end{itemize}

 The red solid line in Fig. \ref{Fig:RelaxationCycles}ab shows that the normal stress difference decreases more at each cycle; for instance, 6 cycles yield a decrease by a factor of 10, apparently without saturating. 
In 3D the procedure is the same, rotating the shearing direction successively along the $x$, $y$ and $z$ axes \cite{Kraynik2003}. 
This procedure is easy to apply to simulations, especially of fully periodic foams.
In experiments, a special set-up should be built: in 2D, it can be a rubber frame  in the spirit of refs. \cite{Abdelkader1999,QuillietPreprint},  if the four corners can be independently displaced.

\begin{figure}[!h]
\setlength{\unitlength}{1cm} 
\centering
\begin{picture}(7,4.5)(0.0,0.0)
\put(-0.7,-0.1){\resizebox{4.5cm}{!}{\includegraphics{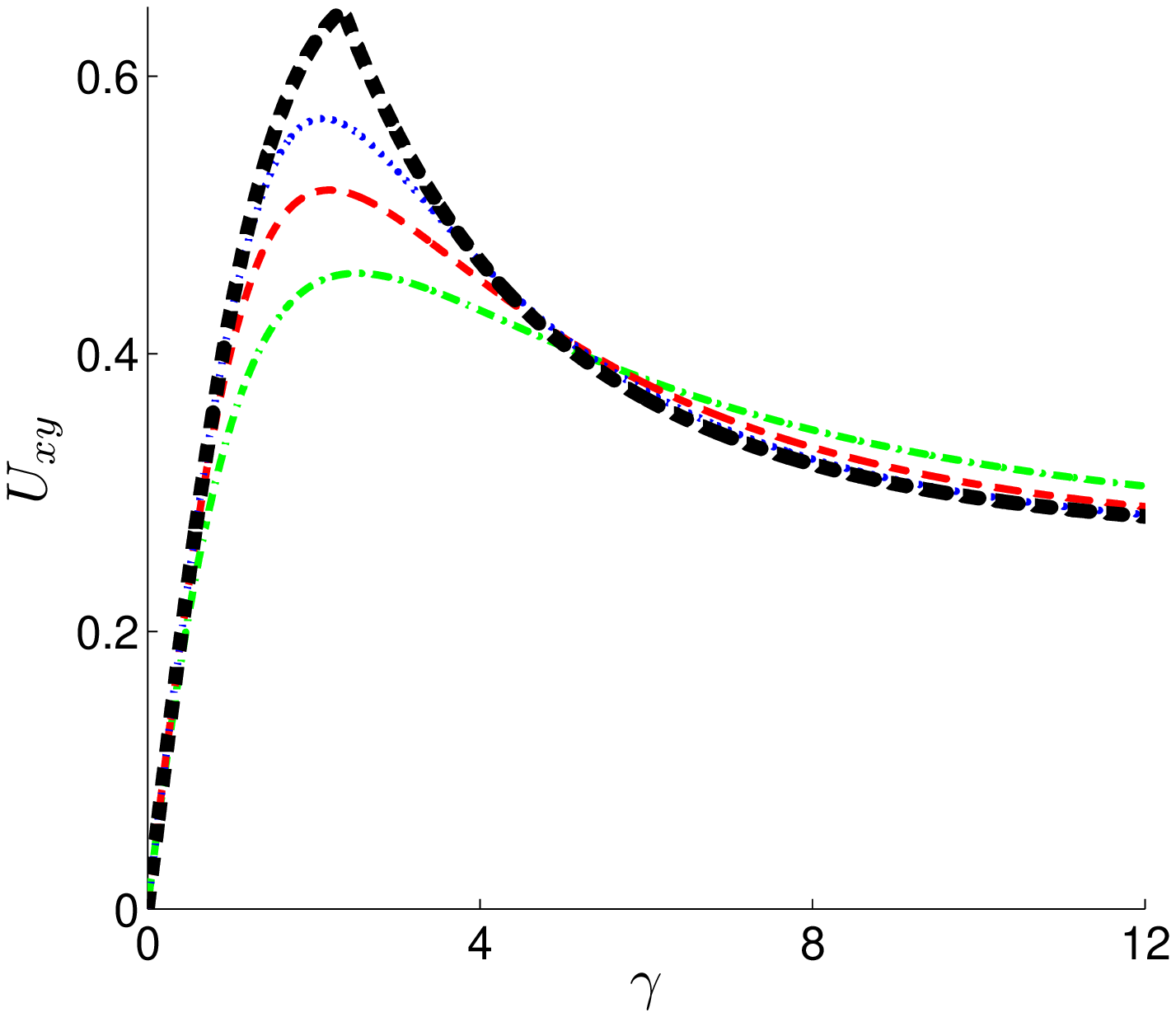}}}
\put(3.5,0){\resizebox{4.5cm}{!}{\includegraphics{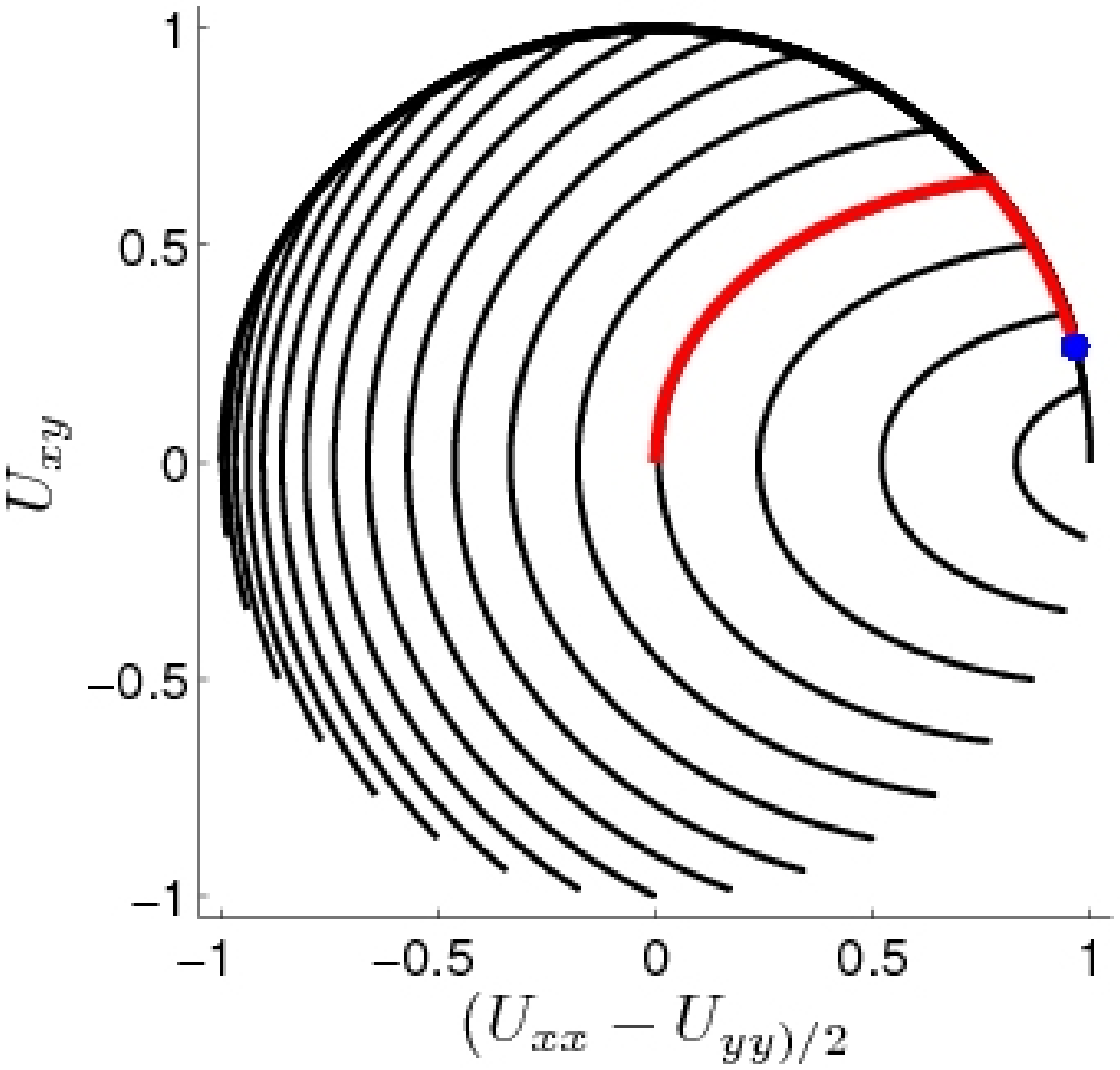}}}
\put(-0.5,0){\large{\textbf{a)}}}
\put(4,0){\large{\textbf{b)}}}
\end{picture}
\caption{Representations of the model for $U_Y = 1$. 
a)
 $U_{xy}$ versus $\gamma$ for the first step and different yield functions, as in Fig. \ref{Fig:Pla-h-n}a. 
  b) Representation of the model for $\dot{\gamma} > 0$ in  \emph{component spaces}
 for $n \rightarrow +\infty$,
  as in Fig. \ref{Fig:ModelCurvesUyPlast}b.
}
 \label{Fig:BigovershootU1}
\end{figure}

\begin{figure}[!h]
\centering
\setlength{\unitlength}{1cm} 
\begin{picture}(8,4.5)(0.0,0.0)
\put(0,-0.5){\resizebox{8cm}{!}{\includegraphics{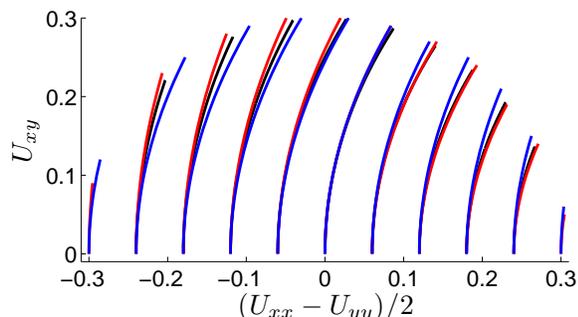}}}
\end{picture}
\caption{Representations of the model for small strains. $U_{xy}$   {\it vs}  $U_{n}=(U_{xx}-U_{yy})/2$. Black: exact model. Blue: first parabolic approximation $U_{n}=U^i_{n}+ U_{xy}^2$. Red:  complete parabolic approximation   (Eq. \ref{Eq:ApproximationComplete}). }
\label{Fig:test}
\end{figure}

\subsection{Materials with low and high $U_Y$}

For practical purposes we plot the reference curves for two limiting types of materials: those with $U_Y$ much higher or much lower than 0.23. 

Fig. \ref{Fig:BigovershootU1} shows the example of $U_Y=1$. The plastic limit corresponds to a small angle $\theta_Y$, resulting in a strong overshoot. 

Fig. \ref{Fig:test} shows that for small strains  in the elastic regime,  all curves can be expressed using a single parameter; for instance, as here, $U^i_{n}$, which is the normal elastic strain $U_{n}=(U_{xx}-U_{yy})/2$ at zero tangential shear ($U_{xy}^i=0$).  A rough parabolic approximation,  and a refined one   (Eq. \ref{Eq:ApproximationComplete}) are plotted here for $U_Y=0.3$. For smaller $U_Y$, this approximation  is good over its whole range of validity (namely the elastic regime), but this range is smaller.

\section{Summary}
\label{sec:Conclusions}

We propose a continuous model of the elasticity and plasticity of disordered, discrete materials such as cellular patterns (for instance liquid foams or emulsions) and assemblies of particles  (for instance colloids). It is based on statistical quantities including (i) the elastic strain $\tensor{U}$, a dimensionless quantity measurable on images, which facilitates the comparison between different experiments or models, and makes apparent the effect of shear on the material's structure; (ii) the yield strain $U$, a classical criterion for the transition between reversible,  elastic and irreversible, plastic regimes; (iii) and the yield function $h(U/U_Y)$, which describes how progressive this transition is, by measuring the relative proportion of elastic and plastic deformation.  They suffice to relate the discrete scale with the collective, global scale. At this global scale, the material behaves as a continuous medium; it is described with tensors such as strain, stress and velocity gradient. We give the  differential equations which predict the elastic and plastic  behaviour. The model is fully tensorial and thus general, in 2D or in 3D. 

We study in detail the case of simple shear. An original representation, suitable for 2D incompressible materials, is introduced to follow the evolution of the material during shear. 

Since $\tensor{U}$ is a tensor, it has an orientation and an amplitude, which both evolve under shear.
It can continuously decrease its amplitude, change direction and increase again its amplitude without ever vanishing (as opposed to a scalar, which can change sign only when it is equal to zero).
Predictions of the model regarding orientation and stretching are plotted. They include a rotation of the structure, which can induce an overshoot of the  shear strain or shear stress (and a smaller, rarer overshoot in normal stress differences) even without overshoot in the elastic strain amplitude. 
This  purely tensorial effect exists if $U_Y$ is at least of order of 0.3. Independently, the shear can also induce a change in the material's structure, sometimes resulting in a (purely scalar) overshoot in  the modulus of the elastic strain. 

The model extends a classical plasticity criterion to disordered media. It can be solved numerically and yields testable predictions. We successfully compare them with carefully converged  quasistatic simulations of shear cycles in 2D foams: the elastic strain increases, saturates and reverses.
 From this comparison between model and simulation we determine   $U_Y$ and estimate  $h$. This method is similar to that which we have used in experiments to extract   $U_Y$  \cite{Cartes,Cheddadi2008}, and a rough estimate of  $h$. We still lack a model to predict $U_Y$ and $h$. Both quantities evolve throughout the simulation, probably due to the evolution of the foam's internal structure, as well as the disorder and fluctuations. In short, the material obeys a continuous description determined by its average properties, while $U_Y$ and $h$ account for the effect at large scale of its fluctuations.

 All quantities involved in the model are directly measurable, as tensors,  in the current state of the material; this includes trapped stresses which we discuss (we also explain how to relax them):
 the history of the sample which led to this current state plays no other direct, explicit role. 
 We explain how and when to use the model   in practice, and provide a set of curves and analytical approximations, including a discussion and an extension of the Poynting relation. At low strain, typically below 0.2, tensorial effects   vanish and an approximate scalar simplification holds.
 
\section*{Acknowledgements}

We thank  C. Quilliet, B. Dollet, S. Attai Talebi and other participants in the Grenoble Foam Mechanics Workshop 2008 for stimulation and useful discussions. We thank K. Brakke for his development and maintenance of the Surface Evolver code. FG thanks Alexandre Kabla for critical reading of the manuscript and recalling the link between overshoot and bistability. SJC thanks the British Council Alliance programme, CNRS and EPSRC (EP/D048397/1, EP/D071127/1) for financial support and UJF for hospitality during the period in which this work was conceived. CR thanks the Alliance programme for having supported one visit to Aberystwyth University, project 15154XB \emph{Foam rheology in two dimensions}.

\appendix

\section{Detailed equations}
\label{sec:DetailedEquations}

\subsection{Notation for tensors}
\label{sec:NotationsForTensors}

We collect here a list of our notation, since the definitions are scattered throughout the text. For a symmetric tensor $\tensor{A}$, we denote by $A_1$ and $A_2$ its eigenvalues, by $A_1+A_2 = A_{xx}+A_{yy}$ its trace, by $A_{xy}=A_{yx}$ its off-diagonal term, by $A_n=(A_{xx}-A_{yy})/2$ half its normal difference, 
by $A  = \sqrt{ \left( A_{xx}-A_{yy} / 2 \right)^2 + U_{xy}^2}$ its amplitude, 
and by $ \left|   \left|   \tensor{A} \right| \right| = A \sqrt{2} $ its euclidian norm defined as 
$ \left|   \left|   \tensor{A} \right| \right|^2 =   \tensor{A} : \tensor{A}$.
The scalar product of two tensors $\tensor{A}$ and $\tensor{B}$ is defined as  $\tensor{A} : \tensor{B} = \sum_{ij} A_{ij} B_{ij}$. 

For a traceless tensor, $A_1=-A_2>0$. If $\theta$ is the angle corresponding to $A_1$, then the point with coordinates $(A \cos \theta, A \sin \theta)$ is a representation of the actual direction of the tensor (\emph{physical space}). Conversely, the point with coordinates $A_{n}=A \cos 2\theta$ and $A_{xy}=A \sin  2\theta$ directly represents the components of the tensor  (\emph{component space}). For a tensor, $\theta$ is defined modulo $\pi$ (and not $2\pi$ as for vectors), so that $2\theta$ has usually more relevance than $\theta$.   
Similarly, for traceless tensors with eigendirections making a relative angle $\phi$, their scalar product is proportional to $\cos 2\phi$. 
This scalar product is maximal when the two eigenvectors of the positive eigenvalues coincide, and minimized when they are perpendicular.  

\subsection{Complete system of equations}
\label{sec:CompleteSystemOfEquations}

We have obtained \cite{RaufasteThese} a complete (closed) set of equations:
\begin{subeqnarray}
\label{Eq:Fundament_ann}
\slabel{Eq:Fundament_1_ann}
 \rho \frac{d}{dt}\vec{v} & = &  \nabla\cdot\left( -p \tensor{Id} + 2 \mu \tensor{U}\right), \\
\slabel{Eq:Fundament_2_ann} 
  \mathtt{div} \; \vec{v} & = & 0,  \\
 \slabel{Eq:Fundament_4_ann}	
   \tensor{U} &  =  & \frac{1}{2} \left(  \log \tensor{M} - \log   \tensor{M}_0 \right), \\
\slabel{Eq:Fundament_3_ann}
  \frac{d}{d t}  \tensor{M}  & = &  \tensor{M}.\tensor{\nabla v} + \tensor{\nabla v}^{t}.\tensor{M} 
- 2 \tensor{P}.\tensor{M}. \nonumber\\
\end{subeqnarray}
  Eq. \ref{Eq:Fundament_1_ann} is the equation of dynamics, equivalent to Navier-Stokes, except that here the viscous stress is assumed to be negligible compared to the elastic stress. 
  Eq. \ref{Eq:Fundament_2_ann} assumes that the flow is  incompressible; this assumption is often valid for foams at small deformation but can be relaxed if needed.
   Eq. \ref{Eq:Fundament_4_ann} defines  the elastic strain from the texture  \cite{Outils}, that is, it assumes that each bubble's  internal degrees of freedom depend on its shape.   
   Eq. \ref{Eq:Fundament_3_ann} is the evolution of the texture, see Eq. \ref{Eq:EvolutionEquation} for the definitions of its terms (transport and source). 
 Here the  plasticity rate $ \tensor{P}$ is predicted according to Eq. 22 in ref. \cite{Cartes}:
\begin{equation}
 \tensor{P} = \frac{1}{2}  \left(\frac{\tensor{U}}{U}:\tensor{\nabla v}_{sym}\right)  {\cal H}\left(\frac{\tensor{U}}{U}:\tensor{\nabla v}_{sym}\right) h\left(\frac{U}{U_{Y}}\right) \frac{\tensor{U}}{U}.
\label{Eq:PlasticMarmottant}
\end{equation}
The meaning of each term is the following. 
The direction of $\tensor{P}$ is set by that of $\tensor{U}$, indicating that the plasticity is opposed to the increase of $\tensor{U}$. 
The amplitude of $\tensor{P}$, that is the rate of plastic rearrangements, is the inverse of a time. It is determined by  the total strain rate  $\tensor{\nabla v}_{sym}$; more precisely, by one component of   $\tensor{\nabla v}_{sym}$, determined by the scalar product with  $\tensor{U}$  (and only if this scalar product is positive, as expressed by the Heaviside function ${\cal H}$). Finally, the amplitude of $\tensor{P}$ depends on the yield criterion, as expressed by $h\left( U/U_{Y}\right)$: the plasticity appears (progressively or abruptly)  when $U$ approaches then exceeds the yield strain.

Eq. \ref{Eq:PlasticMarmottant} is written here by assuming that $\tensor{M}$ and $\tensor{U}$ commute (see Eq. 20 of ref. \cite{Outils}),
which is always the case if $ \tensor{M}_0$ is isotropic. Like Eq. \ref{Eq:Fundament_2_ann}, it assumes that the flow is incompressible, but can be extended to more general cases. It also assumes that the flow is slow: see ref. \cite{MarmottantPinceau} for a discussion of ``quasistatic" flow, and
 \cite{Saramito2007,Cheddadi2008} for the extension to higher velocity. 

The next appendices examine more restrictive cases, that is, additional approximations:
simple shear (\ref{sec:TensorComponents}), 
small strain (\ref{sec:AnnScalarLimit}),
and the purely elastic regime
(\ref{sec:AnalyticalApproximation}).

\subsection{Simple shear}
\label{sec:TensorComponents}

In our geometry, the notation becomes:
$$
\begin{array}{rcl}
\tensor{\nabla v}_{sym}&=& 
\displaystyle \frac{\dot{\gamma}}{2}
\left(
\begin{array}{cc}
0 & 1 \\
1 & 0
\end{array}
\right),
 \\
\tensor{M} &=&
\left(
\begin{array}{cc}
M_{xx} & M_{xy} \\
M_{xy} & M_{yy}
\end{array}
\right),
\\
\tensor{U} &=&
\left(
\begin{array}{cc}
U_{xx} & U_{xy} \\
U_{xy} & U_{yy}
\end{array}
\right),
\\
\tensor{U} : \tensor{\nabla v}_{sym} &=& U_{xy} \dot{\gamma}=U \dot{\gamma} \sin 2\theta
.
\end{array}
$$

Here, due to our conventions, the angle between both tensors is $\phi = \theta - 45^\circ$, hence the term $\cos 2 ( \theta - 45^\circ) =  \sin 2\theta.$
This scalar product is maximal when the two eigenvectors of the positive eigenvalues coincide (which happens for $\theta = 45^\circ$), and minimized when they are perpendicular ($\theta=90^\circ$).

\subsection{Elasto-plastic component equations}
\label{sec:ElastoPlasticComponentEquations}

Under simple shear the advection term is supposed equal to zero and Eq. \ref{Eq:Fundament_3_ann} becomes
\begin{subeqnarray}
\label{Eq:Ann2EquationQuasistatique}
\slabel{Eq:Ann2EquationQuasistatique_1}
\frac{1}{\dot{\gamma}} \partial_{t} M_{xx} & = & 2 M_{xy}   -  \frac{U_{xy}}{U^2} {\cal H}\left(\dot{\gamma} U_{xy} \right) h\left(\frac{U}{U_{Y}}\right) \left[\stackrel{=}{U} .\stackrel{=}{M}\right]_{xx} \nonumber \\
& &  \\
\slabel{Eq:Ann2EquationQuasistatique_2}
\frac{1}{\dot{\gamma}} \partial_{t} M_{yy} &  = &  -  \frac{U_{xy}}{U^2} {\cal H}\left(\dot{\gamma} U_{xy} \right) h\left(\frac{U}{U_{Y}}\right) 
\left[\stackrel{=}{U} .\stackrel{=}{M}\right]_{yy} \nonumber \\
& &  \\
\slabel{Eq:Ann2EquationQuasistatique_3}
\frac{1}{\dot{\gamma}} \partial_{t} M_{xy} & = & M_{yy}  -  \frac{U_{xy}}{U^2} {\cal H}\left(\dot{\gamma} U_{xy} \right) h\left(\frac{U}{U_{Y}}\right) \left[\stackrel{=}{U} .\stackrel{=}{M}\right]_{xy}. \nonumber \\
& & 
\end{subeqnarray}
The elastic regime can be studied by taking the last term of these equations equal to 0 (limit of high $U_Y$). The plastic limit is calculated by taking the left hand sides of these equations equal to 0, $h=1$, and ${\cal H}=1$.

\subsection{Scalar limit}
\label{sec:AnnScalarLimit}
  
In the limit of small strain, $\tensor{U}$ can be linearized:
\begin{equation}
\tensor{U}  =  \frac{1}{2 \lambda_0} (\tensor{M} - \tensor{M}_0),
\label{Eq:Linearized}
\end{equation}
where $\lambda_0$ is the isotropic eigenvalue of $\tensor{M}_0$. In that limit Eq. \ref{Eq:Ann2EquationQuasistatique_3} becomes
\begin{equation}
\label{Eq:LinkTensorScalar}
\frac{1}{\dot{\gamma}} \partial_{t} U_{xy}  =  \left(U_{yy}+\frac{1}{2}\right)  - \frac{1}{2} \left(\frac{U_{xy}}{U}\right)^2 {\cal H}\left(\dot{\gamma} U_{xy} \right) h\left(\frac{U}{U_{Y}}\right).
\end{equation}
Assuming that $\theta$ remains  close to $45^\circ$  leads to
\begin{subeqnarray}
\label{Eq:Ann2Orientation45}
\slabel{Eq:Ann2Orientation45_1}
U_{xx} & = & U_{yy} = 0, \\
\slabel{Eq:Ann2Orientation45_2}
|U_{xy}| &  = &  U, \\
\slabel{Eq:Ann2Orientation45_3}
 \partial_{t} 2 U_{xy} & = & \dot{\gamma}  - \dot{\gamma} {\cal H}\left(\dot{\gamma} U_{xy} \right) h\left(\frac{|U_{xy}|}{U_{Y}}\right).
\end{subeqnarray}
The last equation is identified as the scalar elasto-plastic equation \cite{MarmottantPinceau}, by taking $2 U_{xy}$ as the scalar elastic strain.

\subsection{Analytical approximation at small strain}
\label{sec:AnalyticalApproximation}

In a purely elastic regime, the evolution equation for the texture is
\begin{eqnarray}
\begin{array}{ccc}
M_{yy} & = & M_{yy}^i\\
M_{xy} & = & M_{yy}^i \gamma + M_{xy}^i\\
M_{xx} & = & M_{yy}^i \gamma^2 + 2 M_{xy}^i \gamma + M_{xx}^i .
\end{array}
 \end{eqnarray}
To express all curves analytically, we  choose a single parameter, for instance the elastic strain in a non-sheared state ($U^i_{xy}= M^i_{xy} =0$):
\begin{equation}
M_{n}=M^i_{n}+\frac{M_{yy}^i}{2} \gamma^2,
\end{equation}
\begin{equation}
M_{xy}=M_{yy}^i \gamma,
\end{equation}
which can be rewritten by eliminating $\gamma$:
\begin{equation}
M_{n}=M^i_{n}+\frac{M_{xy}^2}{2 M^i_{yy}}.
\label{Eq:step1}
\end{equation}
There are still two constants left, $M^i_{n}$ and $M^i_{yy}$. To eliminate one of them, we use the fact that the trace of $\tensor{U}$ is almost zero, and thus the determinant of $\tensor{M}$ is almost constant:
\begin{equation}
M^i_{xx} M^i_{yy} = \lambda^2_0,	
\label{determinM}
\end{equation}
or equivalently, using Eq. \ref{Eq:step1}:
\begin{equation}
(2 M^i_{n} + M^i_{yy}) M^i_{yy} = \lambda^2_0	.
\label{determin}
\end{equation}
Solving Eq.  \ref{determin} yields
\begin{equation}
\frac{M^i_{yy}}{\lambda_0} = - \frac{M^i_{n}}{\lambda_0} + \sqrt{\left(\frac{M^i_{n}}{\lambda_0}\right)^2 + 1}
\end{equation}
or equivalently, eliminating $M^i_{yy}$ using Eq.  \ref{determin}: 
\begin{equation}
M_{n}=M^i_{n}+ \left(\sqrt{\left(\frac{M^i_{n}}{\lambda_0}\right)^2 + 1} + \frac{M^i_{n}}{\lambda_0}\right) \frac{M_{xy}^2}{2 \lambda_0}.
\label{M_n}	
\end{equation}
Coming back to $\tensor{U}$ using Eq. \ref{Eq:Linearized}:
\begin{equation}
U_{xy}=\frac{1}{2 \lambda_0} M_{xy},	
 \quad U_{n}=\frac{1}{2 \lambda_0} M_{n}.
\end{equation}
Eq.  \ref{M_n} yields	
a parabolic approximation:
\begin{equation}
U_{n}=U^i_{n}+ \left(\sqrt{\left(\frac{U^i_{n}}{\lambda_0}\right)^2 + 1} + \frac{U^i_{n}}{\lambda_0}\right) U_{xy}^2	.
\label{Eq:ApproximationComplete}
\end{equation}
The parameter which determines each elasticity curve  is the normal strain difference at zero shear (which is thus equal to the amplitude of elastic strain at zero shear). Eq. \ref{Eq:ApproximationComplete} is tested on Fig. \ref{Fig:test} for $U$ up to 0.3. The prefactor of the parabola, {\it i.e.} the bracket in Eq. \ref{Eq:ApproximationComplete}, is exactly 1 if $U^i_{n}=0$: this is the  Poynting relation  \cite{Labiausse2007}  (black curve on Fig.  \ref{Fig:test}, starting from the point $U_{n}=U_{xy}=0$). In fact, even for $U^i_{n} \neq 0$, the bracket in Eq. \ref{Eq:ApproximationComplete} remains close to 1:
as shown in Fig.  \ref{Fig:test}, the  Poynting relation extends even to  an initially anisotropic material.

\end{document}